\documentclass[letterpaper, 12pt]{iopart}
\usepackage{iopams,setstack,graphicx,multirow,hyperref,doi,siunitx, xcolor, ulem}
\hypersetup{
    colorlinks=true,       % false: boxed links; true: colored links
    linkcolor=blue,          % color of internal links (change box color with linkbordercolor)
    citecolor=blue,        % color of links to bibliography
    filecolor=blue,      % color of file links
    urlcolor=blue           % color of external links
}

\graphicspath{{./}{./figs/}}

\begin{document}
\title{Grating magneto-optical traps with complicated level structures}
\author{
Daniel S. Barker$^1$,
Peter K. Elgee$^2$,
Ananya Sitaram$^2$,
%James A. Fedchak$^1$,
Eric B. Norrgard$^1$,
Nikolai N. Klimov$^1$,
Gretchen K. Campbell$^2$,
%Julia Scherschligt$^1$,
Stephen Eckel$^1$}
\address{$^1$ Sensor Science Division, National Institute of Standards and Technology, Gaithersburg, MD 20899, USA}
\address{$^2$ Joint Quantum Institute, University of Maryland, College Park, MD 20742, USA}
\eads{\mailto{daniel.barker@nist.gov}, \mailto{stephen.eckel@nist.gov}}

\begin{abstract}
We study the forces and optical pumping within grating magneto-optical traps (MOTs) operating on transitions with non-trivial level structure.
In contrast to the standard six-beam MOT configuration, rate equation modelling predicts that the asymmetric laser geometry of a grating MOT will produce spin-polarized atomic samples.
Furthermore, the Land\'e \(g\)-factors and total angular momenta of the trapping transition strongly influence both the confinement and equilibrium position of the trap.
Using the intuition gained from the rate equation model, we realize a grating MOT of fermionic \(^{87}\)Sr and observe that it forms closer to the center of the trap's quadrupole magnetic field than its bosonic counterpart.
We also explore the application of grating MOTs to molecule laser cooling, where the rate equations suggest that dual-frequency operation is necessary, but not sufficient, for stable confinement for type-II level structures.
To test our molecule laser cooling models, we create grating MOTs using the \(D_1\) line of \(^7\)Li and see that only two of the four possible six-beam polarization configurations operate in the grating geometry.
Our results will aid the development of portable atom and molecule traps for time keeping, inertial navigation, and precision measurement.
\end{abstract}
%\pacs{???}
\submitto{\NJP}
\maketitle

\section{\label{sec:intro} Introduction}
Magneto-optical traps (MOTs) are the modern workhorse of atomic physics, allowing for the creation of ultra-cold samples of atoms for high-precision spectroscopy~\cite{Zhang2015, Miyake2019}, time keeping~\cite{Hinkley2013, Ohmae2021}, inertial sensing~\cite{Cassella2017, Rudolph2020}, quantum simulation~\cite{Eckel2018b,Subhankar2019}, and quantum computation~\cite{Graham2022, Bluvstein2022}, among others.
The typical MOT configuration uses three orthogonal pairs of red-detuned counter-propagating laser beams in combination with a spherical quadrupole magnetic field to provide both slowing, through the Doppler effect, and spatial confinement, through the Zeeman effect~\cite{Raab1987}.
This six-beam MOT configuration is generally realized in a laboratory, where up to 2~m$^3$ of space is used simply for the optics needed align the laser beams and form the MOT.
In the last two decades, efforts to miniaturize MOTs have led to many different types of MOT geometries that attempt to simplify the optical setup~\cite{Rushton2014, McGilligan2022}, including mirror MOTs~\cite{Reichel1999}, pyramidal MOTs~\cite{Lee1996}, tetrahedral MOTs~\cite{Shimizu1991, Vangeleyn2009}, and photonic-integrated-circuit MOTs~\cite{Isichenko2023, Ropp2023}, and others.

\begin{figure}
  \center
  \includegraphics{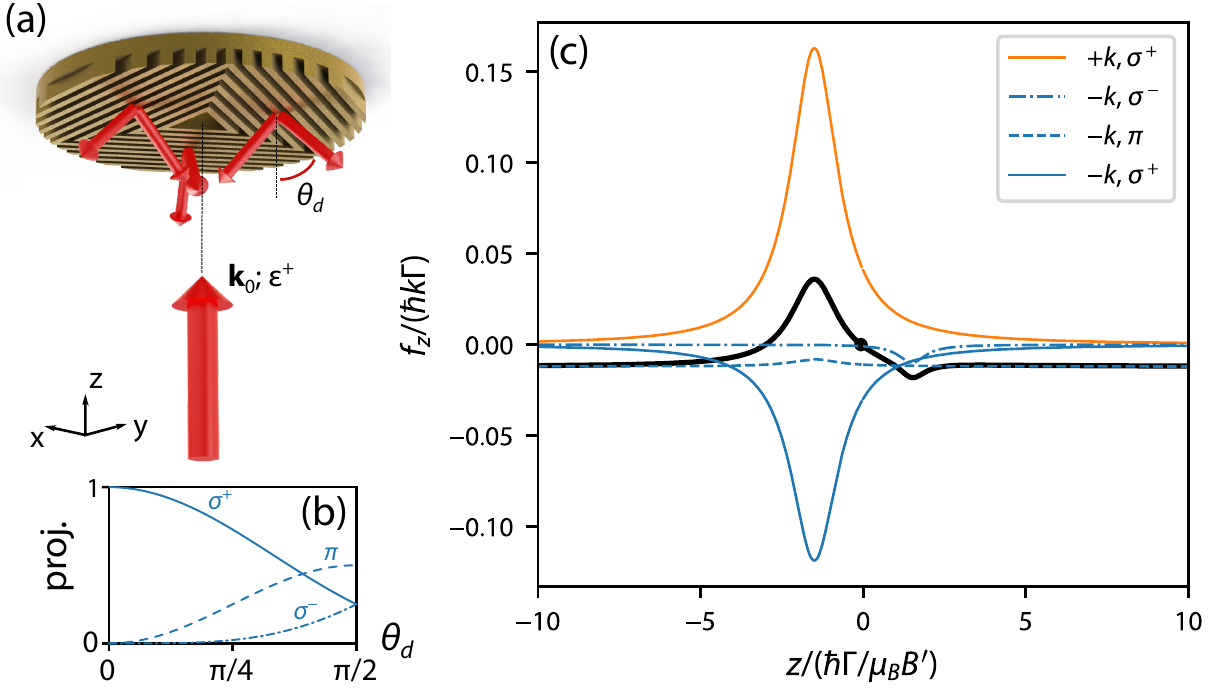}
  \caption{\label{fig:intro}
  Grating magneto-optical trap (gMOT) geometry and forces for a $F=0\rightarrow F'=1$ transition.
  (a) Geometry of a typical gMOT with $N_b=3$.
  The input beam wavevector $\mathbf{k}_{0}$ is directed along $+z$, diffracted beams reflect with diffraction angle $\theta_d$ measured relative to the $-z$ axis.
  The input polarization is circular polarized $\epsilon^+$, which drives $\sigma^+$ transitions when the quantization axis is along $+z$.
  (b) When projected onto the $+z$ axis, the diffracted beams drive different amounts of $\sigma^\pm$ and $\pi$ transitions, depending on the diffraction angle $\theta_d$.
  (c) The resulting force \(f_z\) along $z$ for $x=y=0$ and $v=0$, assuming incident intensity $s_0=1$, $\theta_d=\pi/4$, diffraction efficiency $\eta=0.33$, $\Delta/\Gamma=-3/2$, and excited state Land\'e $g$-factor $g_{F'}=1$ (see Sec.~\ref{sec:theoretical_details}).
  The black curve is the total force, the orange curve is the force from the input beam, and the blue curves are the forces that result from the projected polarization components: $\sigma^+$ (solid), $\pi$ (dashed), and $\sigma^-$ (dashed-dot).
  The black dot on the total force curve shows the axial equilibrium position of the MOT \(z_{\rm eq}\) where $f_z(z_{\rm eq})=0$.
  Note that \(f_z\) is plotted in units of \(\hbar k \Gamma\) and \(z\) is plotted in units of \(\hbar\Gamma/\mu_B B'\) (see Sec.~\ref{sec:theoretical_details}).
  }
\end{figure}

One such configuration is the grating MOT (gMOT)~\cite{Vangeleyn2010, Nshii2013}, shown in figure~\ref{fig:intro}(a).
The gMOT uses a single, incident laser beam in combination with a set of diffraction gratings to produce at least three more beams necessary to form a tetrahedral-like or pyramidal-like MOT.
Tetrahedral gMOTs have the minimum number of beams needed to produce confinement in all three spatial directions~\cite{Vangeleyn2009}.
The beam geometry is similar to that of early tetrahedral MOTs~\cite{Shimizu1991, Lin1991}, which have four beams whose $\hat{k}$ vectors intersect at approximately $109.5^\circ$, instead of the usual right angles for a six-beam MOT.

The greatly simplified setup of gMOTs may prove vital toward MOT miniaturization efforts, as one can imagine simplified chip-scale MOTs with a large grating coupler and custom diffraction grating to produce all necessary laser beams~\cite{McGehee2021, McGilligan2020}.
Grating MOTs are already being incorporated into quantum devices such as microwave clocks~\cite{Elvin2019}, atom interferometers~\cite{Lee2022}, and vacuum sensors~\cite{Ehinger2022}.
Another advantage of having non-right angle intersections is that one of the MOT beams can also be used as a slowing beam to increase the capture velocity of atoms from an oven or other similar directional source.
This advantage was used to realize the first Li and metastable Ne MOTs in the early 1990s~\cite{Lin1991, Shimizu1991}.
Indeed, this advantage was also realized for the first Li and Sr gMOTs, which used an integrated Zeeman slower stage to increase the MOT loading rate~\cite{Barker2019, Sitaram2020}.

The simplicity of gMOTs comes at a cost: gMOTs sacrifice the high degree of symmetry inherent to the standard six-beam MOT~\cite{Raab1987}.
The lack of inversion symmetry in both position and velocity space has several consequences.
Consider the geometry of the gMOT pictured in figure~\ref{fig:intro}(a).
The magnetic field is the standard spherical quadrupole field $\mathbf{B}(\mathbf{r}) = B'z\hat{z} - (B'/2)(x\hat{x}+y\hat{y})$, where $B'$ is axial magnetic field gradient.
The incident beam is axial with normalized wavevector $\hat{k}_{l=0} = \hat{z}$, normalized intensity $s_{l=0} = I_{l=0}/I_{\rm sat}$ (where $I_{\rm sat}$ is the minimum saturation intensity of the atomic transition), and $\hat{\epsilon}_{l=0} = \hat{\epsilon}^+$ right-handed circular polarization.
%Here, $I_{\rm sat}$ is the saturation intensity, specifically defined below.
The diffraction gratings produce \(N_b\) diffracted beams, labeled $l=1\cdots N_b$, that are evenly distributed azimuthally with $\hat{k}_l\cdot(-\hat{z})=\cos\theta_d$, $s_{l} = \eta I_{l=0}\sec\theta_d/I_{\rm sat}$, and the opposite $\hat{\epsilon}_l = \hat{\epsilon}^-$ left-handed circular polarization compared to the incoming beam in the perfectly reflected case.
Here, $\theta_d$ is the first-order diffraction angle and $\eta$ is the first-order diffraction efficiency.
Unless otherwise noted, we will assume all laser beams have uniform intensity with infinite transverse size and that the optical molasses within the gMOT is intensity balanced (\textit{i.e.} \(\eta = 1/N_b\)~\cite{Vangeleyn2010}).
All beams have the same detuning $\Delta_l = \Delta$ from the atomic resonance at \(\mathbf{B}=0\).

At first glance, it appears that the diffracted beams have the incorrect polarization to produce confinement~\cite{Raab1987}.
However, when their polarization is projected onto the magnetic field along $\hat{z}$, they contain light capable of driving $\sigma^\pm$ and $\pi$ transitions [see Fig~\ref{fig:intro}(b)]~\cite{Lee2013}.
The dominant transition for angles $\theta_d\lesssim\pi/3$ is indeed the ``incorrect'' $\sigma^+$ transition, followed by $\pi$ and the ``correct'' $\sigma^-$.
At \(\theta_d = \pi/4\), approximately \(\{73~\%,\, 25~\%,\, 2~\%\}\) of the intensity of the diffracted beams is projected into the \(\{\sigma^+,\, \pi,\, \sigma^-\}\) polarization with respect to the \(+z\) axis.
The resulting forces (derived as in Sec.~\ref{sec:theoretical_details}) are shown in figure~\ref{fig:intro}(c) for the prototypical $F=0\rightarrow F'=1$ atom, where $F$ ($F'$) is the ground (excited) state total angular momentum quantum number.
Experimentally, $F=0\rightarrow F'=1$ transition MOTs are realized with alkaline-earth bosons, which have no electronic angular momentum in their ground state and have no nuclear spin.
The ``incorrect'' $\sigma^+$ polarization component of the diffracted beams, being resonant at the same spatial location as the input beam, reduces the spatial force exerted by the input beam.
For every photon in the input beam, there are $(1+\cos\theta_d)/2$ photons with the same $\sigma^+$ polarization that apply a momentum kick along $\hat{z}$ of $\hbar k \cos\theta_d$ for $\eta= 1/N_b$.
Thus, the ratio of the magnitude of the force from the diffracted beams to the force from the input beam is $(1+\cos\theta_d)\cos\theta_d/2=3/4$ for $\theta_d=\pi/4$, and the net force when the $\sigma^+$ polarization is resonant is positive.
The $\pi$ component of the diffracted beams produces a constant offset force such that the axial equilibrium position of the MOT \(z_{\rm eq}\) is displaced from the quadrupole field zero at \(z=0\) (see figure~\ref{fig:intro}(c)).
%makes the total spatial force cross zero.
%The MOT then forms at this crossing point, shown in Fig~\ref{fig:intro}(c). 

Because of the lack of symmetry, the exact details of the MOT performance -- such as axial trapping frequency, axial equilibrium position, radial trapping frequency, etc. -- are dependent on the relative intensities of input and diffracted beams, including their spatial profiles~\cite{McGilligan2015, Bondza2022}.
Grating MOTs operating on dipole-forbidden transitions must also contend with gravitational shifts to the axial equilibrium position~\cite{Bondza2022}.
Moreover, changing the polarization of the diffracted beams can also have a large effect.
For example, changing the diffracted polarization such that it includes more $\pi$ component, can increase the axial restoring force at the expense of radial trapping~\cite{Imhof2017}.
Thus, gMOTs operate on a careful balance of intensity and control of diffraction angle and diffracted polarization.

Most previous theoretical studies of gMOTs have focused on the prototypical $F=0\rightarrow F'=1$ atom~\cite{Barker2019, McGilligan2015, Imhof2017, Eckel2018, Bondza2022}.
Investigations that considered \(F>0\) concentrated on sub-Doppler cooling and did not include the magnetic field gradient of a MOT~\cite{Lee2013,Barker2022}.
Given that practical devices based on atomic and molecular physics will inevitably use different species depending on the application, a relevant question arises: are there any atomic or even molecular species that can be trapped in a six-beam MOT, but not in a gMOT?
In this article, we theoretically investigate the limitations of gMOTs for real atoms and molecules.
We investigate these questions in the context of a rate equation model, described in Sec.~\ref{sec:theoretical_details}.

We first consider type-I $F\rightarrow F'=F+1$ transitions in Sec.~\ref{sec:typeI}. % in Sec.~\ref{sec:F_to_F}.
In Sec.~\ref{sec:F_to_Fp1}, we show that for $F\neq0$,  gMOTs exhibit position-independent spin polarization due to the ``incorrect'' polarization of the diffracted beams.
The gMOT spin polarization causes a displacement of the axial equilibrium position of the gMOT from the quadrupole field zero that depends of \(F\).
Moreover, as $F$ increases, we find that type-I gMOTs are only stable for certain combinations of ground state \(g_F\) and excited state \(g_{F'}\) Land\'{e} $g$-factors.
When applied to alkaline-earth elements in Sec.~\ref{sec:sryb}, the rate equation model predicts that fermionic alkaline-earth gMOTs form at a different axial equilibrium position than bosonic alkaline-earth gMOTs.
Using the intuition provided by the model, we realize a fermionic strontium gMOT in the apparatus of Ref.~\cite{Sitaram2020} and confirm the rate equation prediction of an \(F\)-dependent axial equilibrium position.
In Sec.~\ref{sec:mitigation}, we examine high \(F\) alkali and alkaline-earth gMOTs that the analysis of Sec.~\ref{sec:F_to_Fp1} indicates will be unstable.
We see that for both alkalis and alkaline-earths the rate equation models suggest that trapping large \(F\) isotopes using transitions with resolved hyperfine structure is possible, but becomes increasingly challenging for large $F$.

We then extend our studies to type-II $F\rightarrow F'=F$ and $F\rightarrow F'=F-1$ transitions in Sec.~\ref{sec:typeII}.
In Sec.~\ref{sec:type-II_simple}, we find that, in contrast to type-I transitions, the spin-polarization effect is devastating: it causes optical pumping into a dark state, eliminating any restoring force.
Stable gMOT operation can be restored using dual-frequency MOT operation~\cite{Tarbutt2015}.
However, in Sec.~\ref{sec:F_to_F} and Sec.~\ref{sec:F_to_Fm1}, we show that dual-frequency gMOTs are only stable for certain combinations of ground and excited state Land\'{e} $g$-factors, limiting the applicability of gMOTs to molecule laser cooling.
Despite rate equation prediction that type-II gMOTs are unstable, we are able to experimentally realize type-II gMOTs on the $D_1$ line of $^7$Li in Sec~\ref{sec:exptype2}.
Extending our model to include the full level structure of the $D_1$ line illuminates the surprising existence of these $D_1$-line gMOTs.
We find that while $D_1$-line gMOTs lack an equilibrium restoring force -- and are therefore not ``traps'' in the traditional sense -- they rapidly recycle atoms through the overlap volume of the MOT laser beams, yielding trap lifetimes and temperatures similar to six-beam $D_1$-line MOTs.
We summarize all of our results and discuss prospects for various quantum devices based on gMOTs in Sec.~\ref{sec:discussion}.
%we show that type-II MOTs formed on the $D_1$ line of $^7$Li, $^{23}$Na, $^{87}$Rb should only form with certain detuning and polarization combinations and with Gaussian beams.
%Finally, we show that alkaline-earth-monofluoride molecules like CaF can also be trapped, with some trade-off in comparison to a standard six-beam MOT.

\section{Theoretical Details}
\label{sec:theoretical_details}
In this work, we use the software package {\tt pylcp} to calculate forces and simulate dynamics, using the rate equations~\cite{Eckel2022}\footnote{The {\tt pylcp} package documentation is available at \url{https://python-laser-cooling-physics.readthedocs.io/en/latest/}.}.
The rate equations neglect coherence between states and, therefore, do not include sub-Doppler forces.
In {\tt pylcp}, the states are labeled by $|n,i\rangle$, where $n$ indexes a manifold of nearly degenerate states (\textit{e.g.} the \(^{2}\)S\(_{1/2}\) hyperfine manifold of an alkali atom) and $i$ indexes the states within manifold $n$ (\textit{e.g.} the Zeeman levels of \(^{2}\)S\(_{1/2}\)).
The fractional population of state $|n,i\rangle$, $N^n_i$, is given by
\begin{eqnarray}
    \label{eq:populations}
    \dot{N}^n_i = & \sum_{m>n,j,l} R^{n\rightarrow m}_{ij,l} (N^m_j - N^n_i) + \sum_{m<n,j,l} R^{m\rightarrow n}_{ji,l} (N^m_j - N^n_i) \nonumber\\
    & + \sum_{m>n} \Gamma^{m\rightarrow n}_{ji} N^m_j - \sum_{m<n}\Gamma^{n\rightarrow m} N^n_i,
\end{eqnarray}
where the first two terms accounts for optical pumping induced by a laser beam with index \(l\), the third term for decays into state $|n,i\rangle$, and the fourth term for decays from state $|n,i\rangle$.
The manifold index $m$ and state index $j$ denote other states $|m,j\rangle$ that are included in the rate equation model.
The manifold indices are ordered by the manifold energy, so that \(\sum_{m>n}\) (\(\sum_{m<n}\)) sums over all manifolds \(m\) with energy larger (smaller) than manifold \(n\).
Every state in manifold $m$ decays to manifold $n$ with an equal rate $\Gamma^{m\rightarrow n}$; the specific decay rate between two states $|m,j\rangle$ and $|n,i\rangle$ is $\Gamma_{ji}^{m\rightarrow n}$.
We note that $\Gamma^{m\rightarrow n}=0$ and $\Gamma_{ji}^{m\rightarrow n}=0$ when the energy of manifold $n$ is greater than the energy of manifold $m$.
The optical pumping rate $R_{ij, l}^{n\rightarrow m}$ due to laser $l$ between states $|n,i\rangle$ and $|m,j\rangle$ is given by
\begin{equation}
    \label{eq:rate_eq:pumping_rate}
    R_{ij,l}^{n\rightarrow m} = \frac{[\Omega^{n\rightarrow m}_{ij,l}]^2/\Gamma^{m\rightarrow n}}{1 + 4[(\Delta_l - (\delta^m_j-\delta^n_i) - \mathbf{k}_l\cdot \mathbf{v})/\Gamma^{m\rightarrow n})]^2},
\end{equation}
where $\delta^n_i$ represents the Zeeman and hyperfine shifts of state $|n,i\rangle$.
The resonant Rabi rate is given by
\begin{equation}
	\label{eq:rabi_rate_rate_eq}
	\Omega_{ij,l}^{n\rightarrow m} = \frac{\Gamma^{m\rightarrow n}}{2} (\mathbf{d}_{ij}^{nm}\cdot \boldsymbol{\epsilon}'_l) \sqrt{2s_l(\mathbf{r},t)},
\end{equation}
where $\mathbf{d}_{ij}^{nm}$ is the dipole matrix element between states $|n,i\rangle$ and $|m,j\rangle$, $\boldsymbol{\epsilon}'_l$ is the polarization of the laser $l$, and $s_l$ is the intensity of laser $l$ measured in terms of the saturation intensity.
The polarization of the light $\boldsymbol{\epsilon}'_l$ decomposes into the usual $\sigma^\pm$ and $\pi$ components when projected onto the local quantization axis, typically defined by the magnetic field.
To determine equilibrium forces in steady state, we first construct the rate equations in terms of a matrix equation with the populations as vectors and then solve for the equilibrium state using singular vector decomposition.
The total force on the atom due to the laser beams is then given by
\begin{equation}
    \label{eq:force}
	\mathbf{f} = \sum_{l} \frac{\hbar \mathbf{k}_l}{2}\sum_{n, i}\sum_{m>n, j} R_{ij,l}^{n\rightarrow m}(N^m_j - N^n_i).
\end{equation}

Throughout the paper, we consider the input beam as being circularly polarized.
We denote left/right circular polarization, defined relative to the laser $\mathbf{k}$ vector, as $\epsilon^{\pm}$.
We always assume a circularly polarized input beam, $\epsilon^+$ unless otherwise noted, but we consider a more general polarization rotation upon reflection in Sec.~\ref{sec:sryb}.
As our basis, we use the common perpendicular ($s$) and parallel ($p$) components relative to the plane of reflection.
A Poincar\'{e} sphere depicts the polarization of the diffracted light, with $s$ and $p$ linear polarization on the equator and the resulting circular polarization $\epsilon^\pm$ on the poles.
If the reflector were perfect, the reflected polarization would be $\epsilon^{-}$ (with $\epsilon^+$ input), and thus we choose $\epsilon^{-}$ to be the north pole.
When projected onto the quantization axis, $+z$ unless otherwise specified, we can then decompose the polarization into $\sigma^\pm$ and $\pi$ components.  

When solving for the equations of motion, we choose to work in a system of units where time is measured in $1/\Gamma$ and position is measured in terms of $1/k$.
Velocity is thus measured in units of the Doppler velocity $v_D = \Gamma/k$; and forces in terms of $\hbar k \Gamma$.
However, when discussing equilibrium forces, we  prefer to measure distances in units of $\hbar\Gamma/\mu_B B'$, where \(\hbar\) is the reduced Planck constant and \(\mu_B\) is the Bohr magneton.

\section{Type-I MOTs}
\label{sec:typeI}
In this section, we build on the textbook $F=0\rightarrow F'=1$ scenario described in Sec.~\ref{sec:intro} to expand our discussion to the more general case of gMOTs operating on $F\rightarrow F'=F+1$ transitions (``type-I MOTs'').
In general, we shall see that pure input polarization with ideal reflection ($\epsilon_0=\epsilon^+$ and $\epsilon_{l>0} = \epsilon^-$) biases the optical pumping to push atoms into the $m_F>0$ states, causing different degrees of problems depending on the nature of the level structure.

\subsection{Basic \texorpdfstring{$F \rightarrow F'=F+1$}{F to F'=F+1} transitions}

\label{sec:F_to_Fp1}
\begin{figure}
  \center
  \includegraphics{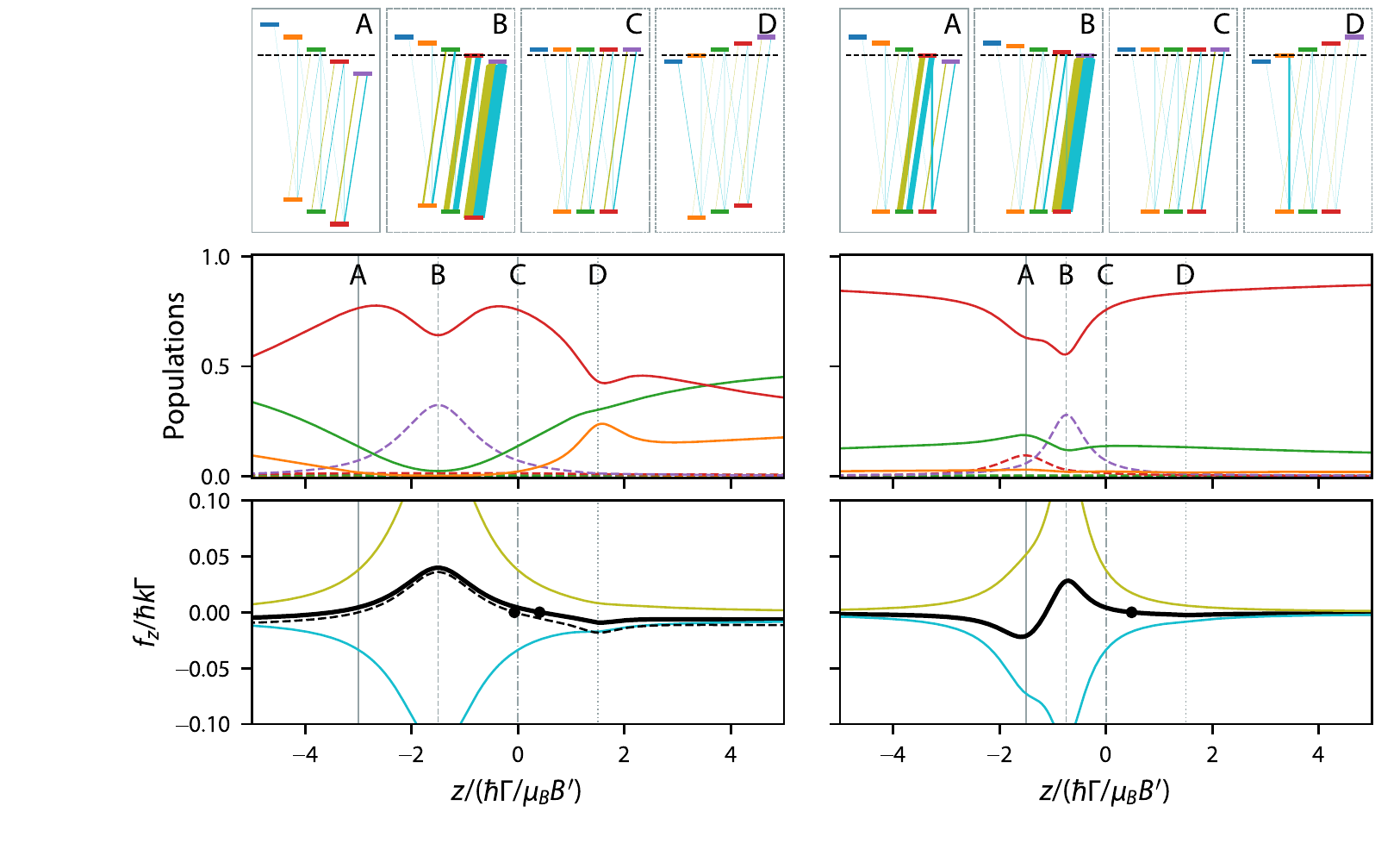}
  \caption{\label{fig:f_to_fp1_level}
  Optical pumping (top row), state populations (middle row), and forces (bottom row) along the $z$ direction in $F=1\rightarrow F'=2$ gMOTs with $\Delta/\Gamma = -3/2$, $s_0=1$, $N_b=3$, $\eta=0.33$, $\theta_d=\pi/4$, $g_{F'} = 1$, and $g_F=1$ (left) or $g_F=0$ (right).
  In all subplots, $x=y=0$ and $v=0$.
  The labels (A--D) of the level diagrams and optical pumping strengths in the top row denote the $z$ positions at which they were calculated through the correspondingly labelled vertical lines in the lower plots.
  The thickness of the olive lines denotes pumping rates for the incident laser beam and the thickness of the cyan lines indicates pumping rates for diffracted laser beams.
  The populations of the ground (solid) and excited (dashed) states in the middle row are colored to match the $m_F$ color coding of the level diagrams.
  Total force in the bottom row plots is the black curve, which is a sum of the force from the incident beam (olive) and the diffracted beams (cyan).
  The dashed black curve shows the force for a $F=0\rightarrow F'=1$ atom [as in figure~\ref{fig:intro}(c)].
  Black dots on the total force curves show the location where the MOT will form ($f_z=0$).
  }
\end{figure}

Figure~\ref{fig:f_to_fp1_level} shows the optical pumping, state populations, and axial forces as a function of position $z$ for a gMOT formed on $F=1\rightarrow F'=2$ with $g_{F'}=1$ and $g_F=g_{F'}$ (left column) or $g_F=0$ (right column).
The former case, $g_F=g_{F'}$ is an ideal case where all transitions of the same type--$\pi$, $\sigma^+$, and $\sigma^-$--are degenerate. 
We are aware of no atom that perfectly realizes this case, but it closely matches the situation in alkali atoms and highly magnetic species like Cr, Dy, and Er.
The latter case, $g_F=0$ corresponds closely to that of alkaline-earth atoms like Ca, Sr and Yb.

Despite the different level structures, the equilibrium populations of the ground state as a function of position $z$ show similar behavior.
For any $g_F$ at $z=0$, all transitions are degenerate and the dominant polarization from both diffracted and incident beams is $\sigma^+$, causing a large population in $|m_F = F\rangle$.
Extending to all $z$, the atom has a non-zero spin projection $\langle F_z\rangle = \sum_{m_F} m_F N_{m_F}$ along $+\hat{z}$.
This spin polarization has markedly different effects, depending on the level structure.

Let us first consider the case where $g_F=g_{F'}$, which is closely analogous to the $F=0\rightarrow F'=1$ case discussed in Sec.~\ref{sec:intro}.
Indeed, comparison of the total force for $F=0\rightarrow F'=1$ [as in figure~\ref{fig:intro}(c)] to $F=1\rightarrow F'=2$ (figure~\ref{fig:f_to_fp1_level} left column, bottom row) when $g_{F'}=g_F$ shows a nearly identical dependence on position $z$.
Two resonances appear at $z = \pm \hbar \Delta/\mu_B B'$, when all the $\sigma^\mp$ transitions simultaneously come into resonance (vertical lines B and D in the left column of figure~\ref{fig:f_to_fp1_level}).
Compared to the $F=0\rightarrow F'=1$ case, the magnitude of force due to the \(\sigma^+\) resonance is mostly unchanged, while the magnitude of force due to the \(\sigma^-\) resonance is reduced.
The reduction in \(\sigma^-\) force occurs because the population is mostly in \(|m_F=0\rangle\) and \(|m_F=1\rangle\), which have reduced Clebsch-Gordan coefficients for \(\sigma^-\) transitions compared to the $F=0\rightarrow F'=1$ case.
The \(\sigma^\pm\) resonances sit atop a nearly spatially-independent force, which is generated by the spatially independent $\pi$ transitions.
Compared to figure~\ref{fig:intro}(c), the magnitude of this spatially-independent force is reduced slightly due to the reduced Clebsch-Gordan coefficients for $\pi$ transitions in a $F=1\rightarrow F'=2$ atom and varies slightly as the populations in the various $|m_F\rangle$ levels shift with position.
The net effect is to shift the axial equilibrium position of the MOT to more positive $z$.

Contrast the $g_F=g_{F'}$ case with the case of $g_F=0$, shown in the right panels of figure~\ref{fig:f_to_fp1_level}, where the spin polarization has a more deleterious effect.
The only magnetic field independent transitions in this case are the $|m_F=\pm1\rangle\rightarrow |m_{F'} = 0\rangle$, driven by $\sigma^{\mp}$ respectively, and $|m_F=0\rangle\rightarrow |m_{F'}=0\rangle$, driven by $\pi$.
Because most of the population is in $|m_F=+1\rangle$ and the spatially-independent transition $|m_{F}=1\rangle\rightarrow |m_{F'}=0\rangle$ is driven by the even smaller $\sigma^-$ component of the diffracted beams, the magnitude of the spatially-independent force is reduced even further compared to the $g_F=g_{F'}$ case.

To further explore the difficulties of the \(g_F=0\) case, consider the resonances in \(f_z\) in the lower right of figure~\ref{fig:f_to_fp1_level}.
At $z= \hbar \Delta/2\mu_B B' = - (3/4) \hbar \Gamma/\mu_B B'$ (vertical line B in the right column of figure~\ref{fig:f_to_fp1_level}), the $|m_F=1\rangle\rightarrow |m_{F'}=2\rangle$ transition comes into resonance, scattering photons from both the incident and diffracted beams, with both more photons and more force per photon being delivered by the incident beam, as described in Sec.~\ref{sec:intro}.
At $z= \hbar \Delta/\mu_B B = - (3/2) \hbar \Gamma/\mu_B B'$ (vertical line A in the right column of figure~\ref{fig:f_to_fp1_level}), the $|m_F=1\rangle\rightarrow |m_{F'}=1\rangle$ $\pi$-polarized transition driven by the diffracted beams comes into resonance, resulting in a negative force.
At other positions where resonances should be observed, like $z=-\hbar \Delta/\mu_B B'= (3/2) \hbar \Gamma/\mu_B B'$ (vertical line D in the right column of figure~\ref{fig:f_to_fp1_level}), where $|m_{F}=-1\rangle\rightarrow |m_{F'}=-1\rangle$ and $|m_{F}=0\rangle\rightarrow |m_{F'}=-1\rangle$ both become resonant, there is not enough population in $|m_F=-1\rangle$ or $|m_F=0\rangle$ sublevels to have any consequential effect on the force. 
As a result, the axial equilibrium position is displaced to larger $z$ compared to the $g_{F'}=g_F$ case, which reduces both the longitudinal and transverse trapping frequencies.
As $F$ becomes larger, the axial equilibrium position displacement becomes more pronounced, such that gMOTs for $F>2$ when $g_F=0$ do not appear feasible.
We further explore the \(F\) dependence of \(z_{\rm eq}\) in Sec.~\ref{sec:sryb} for experimentally realizable gMOTs of Yb and Sr.

\begin{figure}
    \centering
    \includegraphics{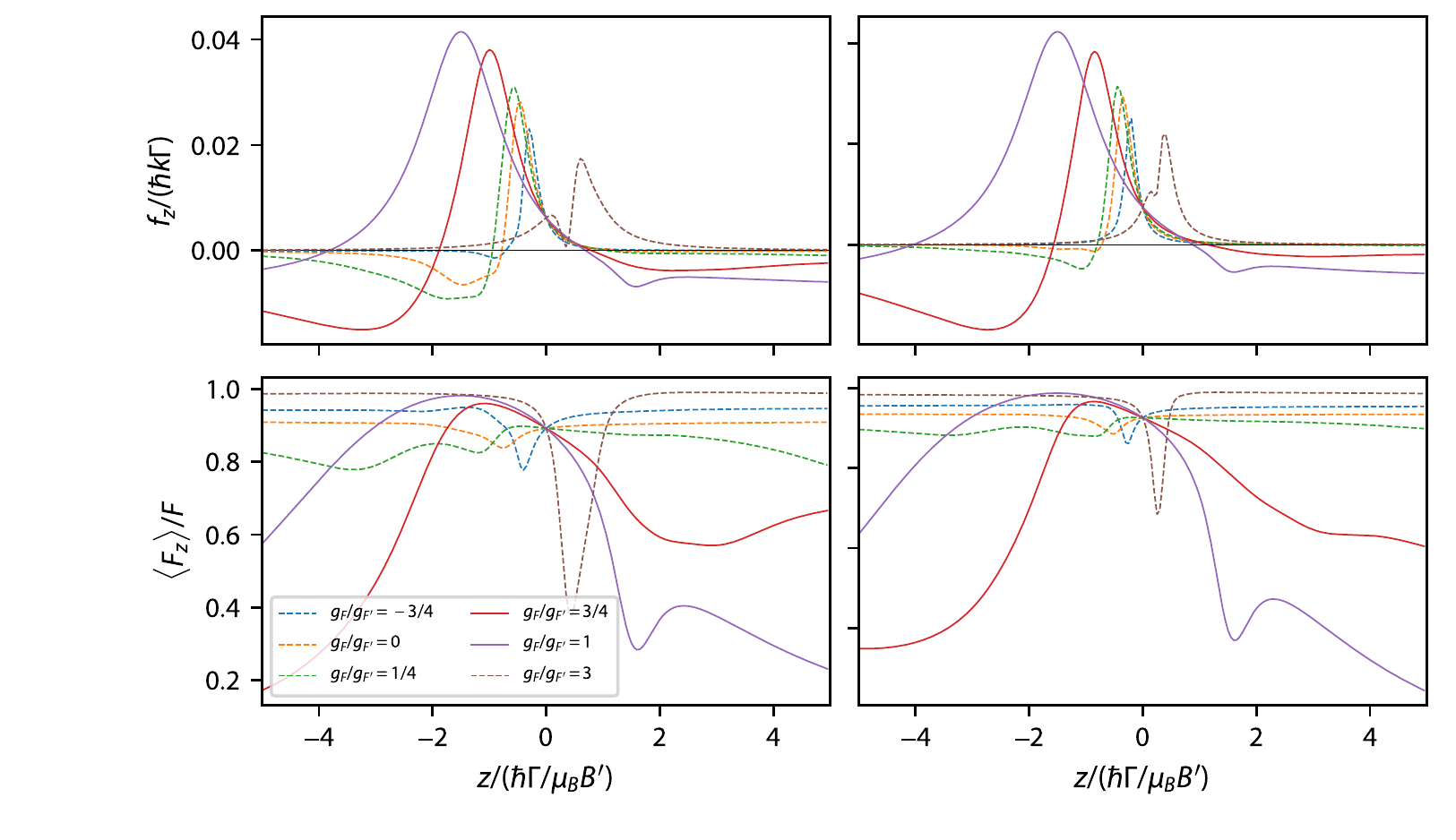}
    \caption{Axial force $f_z$ (top row) and spin polarizations (bottom row) for $F=2$ (left column) and $F=4$ (right column) as a function of $z$ in a gMOT with $\Delta/\Gamma = -3/2$, $s_0=1$, $N_b=3$, $\eta=0.33$, $\theta_d=\pi/4$, and $g_{F'} = 1/F'$.
    In all subplots, $x=y=0$ and $v=0$.
    Solid (dashed) curves represent ground state $g$ factors that do (do not) fulfill Eq.~\ref{eq:stabMOT}.}
    \label{fig:F_to_Fp1_gfactors}
\end{figure}

The contrast between $g_F=0$ and $g_F=g_{F'}$ is reminiscent of the contrast between alkali and alkaline-earth MOTs explored in Ref.~\cite{Mukaiyama2003}.
Namely, we should expect that gMOTs will work if the level structure is analogous to that of the $F=0\rightarrow F'=1$ atom: when all $\sigma^+$ transition energies have a positive slope with magnetic field and all $\sigma^-$ transition energies have a negative slope with magnetic field.
This requirement on the slope of the Zeeman shifts is fulfilled when
\begin{equation}
    \label{eq:stabMOT}
    \frac{F-1}{F}<\frac{g_F}{g_{F'}}<\frac{F+1}{F},
\end{equation}
as shown in Ref.~\cite{Mukaiyama2003}. 
Alkali atom MOTs have \(g_F=1/F\) and \(g_{F'}=2/F'\) for \(|^{2}S_{1/2},F=I+1/2 \rangle \rightarrow\, |^{2}P_{3/2}, F'=F+1\rangle\) transitions, where $I$ is the nuclear angular momentum quantum number.
Thus, Eq.~\ref{eq:stabMOT} is fulfilled for alkalis with $F<4$ (all long-lived alkalis except $^{40}$K and $^{133}$Cs).
Demanding that Eq.~\ref{eq:stabMOT} be satisfied partially alleviates the difficulty of trapping high \(F\) species in a gMOT, as shown in figure~\ref{fig:F_to_Fp1_gfactors}.
For the values of $g_F/g_{F'}$ shown, those that satisfy Eq.~\ref{eq:stabMOT} form stable MOTs for $F=2$ and $F=4$.
Of the displayed $g_F/g_{F'}$ values that violate Eq.~\ref{eq:stabMOT}, only $g_F/g_{F'}=1/4$ forms a stable MOT when $F=2$.
Satisfying Eq.~\ref{eq:stabMOT} appears to be sufficient, but not necessary, for gMOT stability.
We explore the viability of gMOTs for real atoms with level structures that violate Eq.~\ref{eq:stabMOT} in Sec.~\ref{sec:sryb} and Sec.~\ref{sec:mitigation}.

\subsection{Real cases: strontium and ytterbium gMOTs}
\label{sec:sryb}

In our prior experimental investigation of strontium confinement in a gMOT~\cite{Sitaram2020}, we were unable to observe trapping of the fermionic isotope \(^{87}\)Sr. 
Fermionic alkaline-earth atom  have $F=I$ in the ground state and therefore exhibit $g_F\approx 0$ because the ground state magnetic moment is determined by the nuclear magneton.
Our simulations in Sec.~\ref{sec:F_to_Fp1} thus suggest that high \(F\) alkaline-earth fermions, such as \(^{87}\)Sr ($F=9/2$), may not be trappable in a gMOT.
Such a conclusion would not be surprising, since six-beam \(^{87}\)Sr MOTs operating on the narrow \(^{1}S_0\,\rightarrow\,^{3}P_1\) intercombination transition are known to exhibit very short trap lifetimes~\cite{Mukaiyama2003}.
However, the six-beam \(^{1}S_0\,\rightarrow\,^{3}P_1\) MOT lifetime can be extended close to the vacuum limit using sawtooth wave adiabatic passage~\cite{Muniz2018, Snigirev2019} or a secondary ``stirring'' laser to rapidly randomize the ground state populations~\cite{Mukaiyama2003, Boyd2007, DeSalvo2010}.
Alkaline-earth MOTs operating on broad \(^{1}S_0\,\rightarrow\,^{1}P_1\) transitions, such as our gMOT in Ref.~\cite{Sitaram2020}, also have \(g_F\approx0\).
In the standard six-beam geometry, off-resonant excitation of \(F\rightarrow F'=F\) and \(F\rightarrow F'=F-1\) transitions is sufficient to extend the MOT lifetime (\textit{i.e.} broad transition MOTs are ``self-stirring'').

Although the axial spin polarization inherent to a gMOT's tetrahedral beam arrangement disturbs the self-stirring dynamics (see figure~\ref{fig:f_to_fp1_level} and figure~\ref{fig:F_to_Fp1_gfactors}), a long-lived trap for fermionic alkaline-earth isotopes may still be possible if we consider the full hyperfine structure of the excited state.
Because the trap instability is dynamic, leaks in the MOT may not be apparent in plots of the equilibrium MOT forces (see Sec.~\ref{sec:F_to_Fp1}).
We therefore search for instability in fermionic \(^{1}S_0\,\rightarrow\,^{1}P_1\) MOTs by numerically calculating the MOT escape velocity \(v_{\text{esc}}\) using the rate equations.
We model the MOT laser beams with Gaussian spatial modes that are truncated to match the \(1.1~\si{\centi\meter}\) radius patterned area of the grating chip~\cite{Barker2019,Sitaram2020}.
Using the {\tt pylcp} package~\cite{Eckel2022}, we construct the rate equations with parameters that mimic the experimental conditions~\cite{Sitaram2020}: \(s_0 = 1\), \(\Delta/\Gamma = -1\), incident laser beam \(1/e^2\) radius \(w_0=1.2~\si{\centi\meter}\), \(N_b = 3\), \(\eta = 0.33\), and \(B'=6~\si{\milli\tesla\per\centi\meter}\).
Atomic properties for the rate equation model are taken from Refs.~\cite{Baumann1966, Kluge1974, Berends1992, Nagel2005, Yasuda2006, Kleinert2016}.
An escape velocity calculation initializes with an atom with equilibrated state populations at the equilibrium position of the MOT, which may not correspond to the quadrupole field zero~\cite{McGilligan2015}.
The atom has an initial trial velocity \(v_t(\theta, \phi)\) directed radially outward from the trap equilibrium position with azimuthal angle \(\phi\) and polar angle \(\theta\) (defined with respect to the coordinate system in figure~\ref{fig:intro}(a)).
We numerically integrate the rate equations and the classical motion of the atom as it moves through the MOT, neglecting stochastic momentum kicks due to spontaneous emission (\textit{i.e.} we compute the average atomic trajectory).
If the atom remains inside, or returns to, the volume where all MOT laser beams overlap within five periods of the smallest trap frequency (or \(26~\si{\milli\second}\) if that is shorter), then \(v_t(\theta, \phi)\leq v_{\text{esc}}(\theta, \phi)\).
We vary \(v_t(\theta, \phi)\) following a binary search pattern in a velocity space spanned by \(k_B\times 5~\si{\milli\kelvin}\leq m v^2_t(\theta, \phi)/2 \leq k_B\times 20~\si{\kelvin}\) -- where \(m\) is the atomic mass and \(k_B\) is Boltzmann's constant -- and take \(v_{\text{esc}}(\theta, \phi)\) equal to the maximum trial velocity that did not escape the MOT.

\begin{figure}
    \center
    \includegraphics[width=\linewidth]{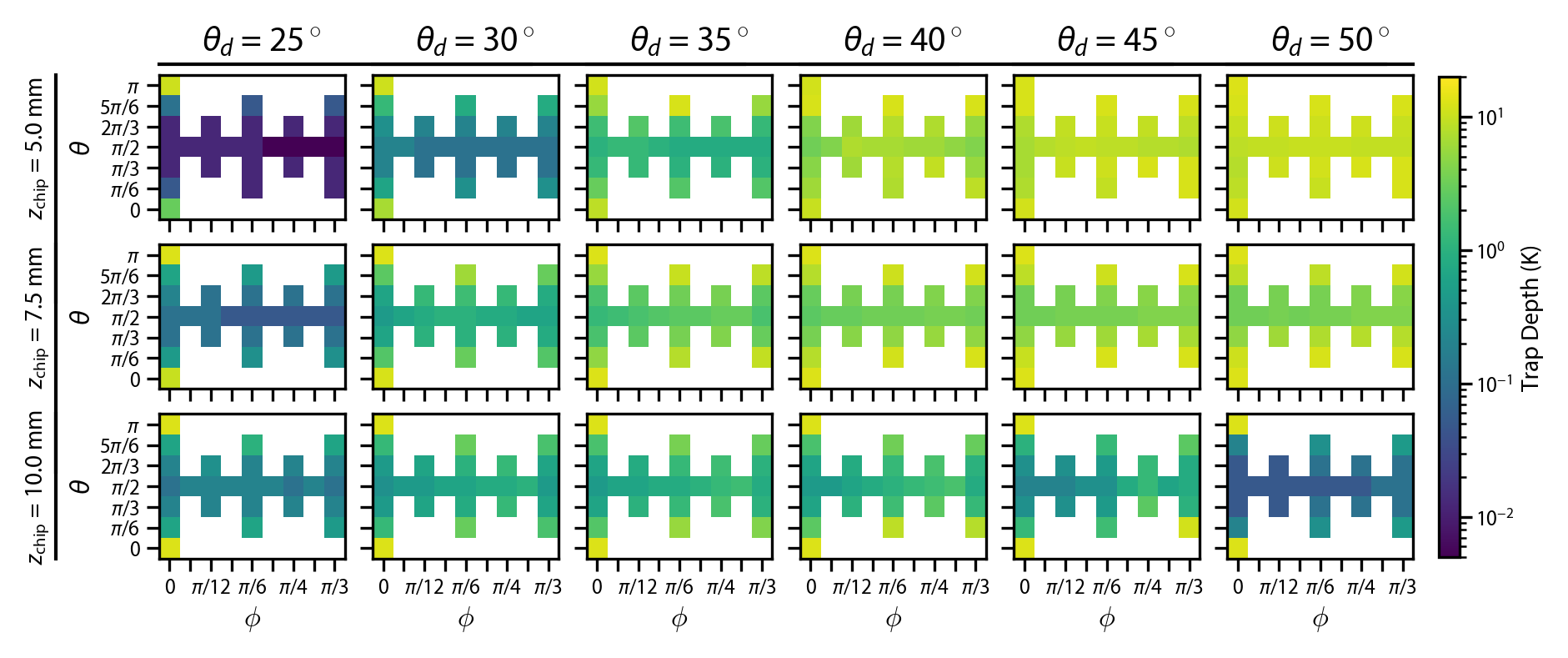}
    \caption{\label{fig:depth_map} \(^{88}\)Sr trap depth maps for various grating chip diffraction angles \(\theta_d\) (increasing by column from left to right) and chip offsets \(z_\text{chip}\) (increasing by row from top to bottom).
    Each map shows the trap depth \(U_d(\theta, \phi) = m v^2_{\text{esc}}(\theta, \phi)/2\) as a function of the azimuthal angle \(\phi\in [0, \pi/3]\) and polar angle \(\theta\in [0, \pi]\) of the initial atomic velocity.
    White regions of each map indicate combinations of \(\phi\) and \(\theta\) for which the trap depth was not computed.
    }
\end{figure}

We produce maps of the escape velocity \(v_{\text{esc}}(\theta, \phi)\) as a function of two experimental parameters: the diffraction angle \(\theta_d\) and the distance from the quadrupole field zero to the surface of the diffraction grating chip \(z_\text{chip}\), which we refer to as the ``chip offset''.
To minimize integration time for each map, we exploit the six-fold symmetry of our diffraction grating chips and quasi-uniformly sample the resulting \(2\pi/3\) solid angle.
Figure~\ref{fig:depth_map} shows example maps of the escape velocity, colored according to the equivalent trap depth \(U_d(\theta, \phi) = m v^2_{\text{esc}}(\theta, \phi)/2\), for \(^{88}\)Sr.
The trap depth is largest for velocities directed parallel or antiparallel to the gMOT's input laser beam.
The difference between the maps in figure~\ref{fig:depth_map} and those we reported in Ref.~\cite{Eckel2018} for \(F\rightarrow F'=F+1\) transitions is expected because Ref.~\cite{Eckel2018} uses a heuristic model for laser cooling that does not include local variation of the transition saturation~\cite{Eckel2022}.
The average trap depth of the gMOT reaches a local maximum when the trap equilibrium position is farthest from the edge of the laser beam overlap volume, which occurs at small \(z_\text{chip}\) for large \(\theta_d\) and at large \(z_\text{chip}\) for small \(\theta_d\).

\begin{figure}
    \center
    \includegraphics[width=\linewidth]{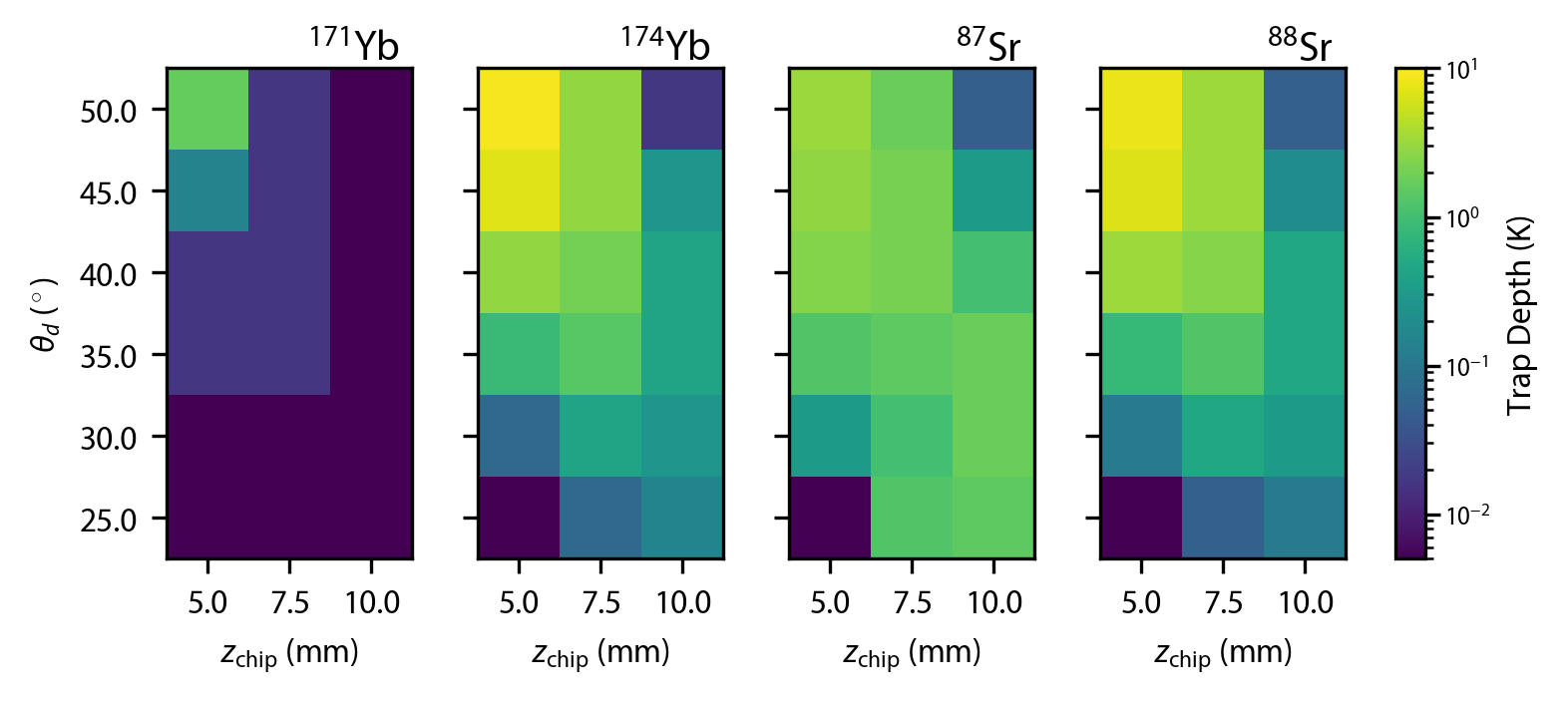}
    \caption{\label{fig:mindepth} Minimum trap depth \(U_d(\theta, \phi)\) as a function of diffraction angle \(\theta_d\) and chip offset \(z_\text{chip}\) for \(^{171}\)Yb (far left), \(^{174}\)Yb (left middle), \(^{87}\)Sr (right middle), and \(^{88}\)Sr (right).
    For all subplots, \(s_0 = 1\), \(\Delta/\Gamma = -1\), \(w_0=1.2~\si{\centi\meter}\), \(N_b = 3\), \(\eta = 0.33\), and \(B'=6~\si{\milli\tesla\per\centi\meter}\).
    For fermionic isotopes, \(\Delta\) is relative to the \(F\rightarrow F'=F+1\) transition.
    }
\end{figure}

In addition to the \(^{88}\)Sr escape velocity maps shown in figure~\ref{fig:depth_map}, we produce escape velocity maps over the same ranges of \(\theta_d\) and \(z_\text{chip}\) for \(^{87}\)Sr ($F=9/2$), \(^{174}\)Yb, and \(^{171}\)Yb ($F=1/2$).
We plot the minimum \(U_d(\theta, \phi)\) for each isotope as a function of \(\theta_d\) and \(z_\text{chip}\) in figure~\ref{fig:mindepth}.
The axial equilibrium positions of the gMOTs (not shown in figure~\ref{fig:mindepth}) increase with both \(\theta_d\) and \(z_\text{chip}\).
The variation in the minimum trap depth across isotopes has three particularly interesting features.
First, the trap depth minima for \(^{174}\)Yb and \(^{88}\)Sr are nearly indistinguishable, as expected given the similar lifetimes of the Yb and Sr \(^{1}P_1\) excited states (approximately \(5.5~\si{\nano\second}\) and \(5.2~\si{\nano\second}\), respectively~\cite{Baumann1966, Nagel2005, Yasuda2006}).
%we might expect given the similar performance of Yb and Sr \(^{1}S_0\,\rightarrow\,^{1}P_1\) MOTs with conventional 6-beam geometries~\cite{Maruyama2003, Xu2003a}.
Second, for \(^{171}\)Yb, the minimum \(U_d(\theta, \phi)\le k_B\times 5~\si{\milli\kelvin}\) (\textit{i.e.} it is consistent with zero given the span of our binary search) over a wide range of diffraction angles and chip offsets.
For all \(\theta_d\) and \(z_\text{chip}\) combinations that we explored, the \(^{171}\)Yb gMOT has transverse trapping frequencies \(\omega_x = \omega_y = 0\).
The lack of transverse confinement suggests that even when the minimum trap depth is greater than zero, \(^{171}\)Yb gMOTs using our parameters will act as atom recyclers -- that keep atoms on looped trajectories through the beam overlap volume -- rather than true MOTs (see Sec.~\ref{sec:exptype2}) and will have short lifetimes.
We will explore and suggest remedies for the lack of transverse confinement in Sec.~\ref{sec:mitigation}.
Third, the minimum trap depths for \(^{88}\)Sr and \(^{87}\)Sr as a function of \(\theta_d\) and \(z_\text{chip}\) agree qualitatively.
The axial spin polarization of gMOTs does not disturb the self-stirring dynamics enough to create leaks in the trap.

\begin{figure}
    \center
    \includegraphics[width=4.5in]{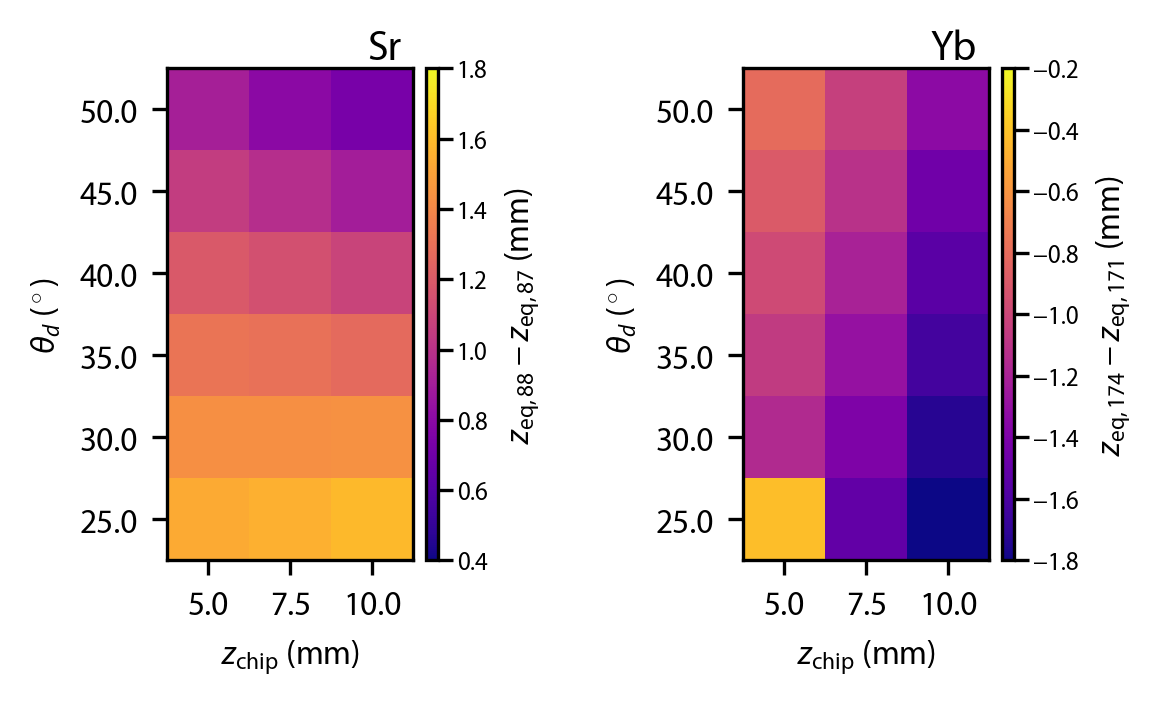}
    \caption{\label{fig:trapshift} Distance between boson and fermion MOT axial equilibrium positions as a function of diffraction angle and chip offset.
    Left: difference in axial equilibrium position of \(^{88}\)Sr and \(^{87}\)Sr gMOTs \(z_{{\rm eq},88}-z_{{\rm eq},87}\).
    Right: difference in axial equilibrium position of \(^{174}\)Yb and \(^{171}\)Yb gMOTs \(z_{{\rm eq},174}-z_{{\rm eq},171}\).
    As in figure~\ref{fig:trapshift} and for both subplots, \(s_0 = 1\), \(\Delta/\Gamma = -1\), \(w_0=1.2~\si{\centi\meter}\), \(N_b = 3\), \(\eta = 0.33\), and \(B'=6~\si{\milli\tesla\per\centi\meter}\).
    For fermionic isotopes, \(\Delta\) is relative to the \(F\rightarrow F'=F+1\) transition.
    }
\end{figure}

\begin{figure}
    \center
    \includegraphics[width=4.5in]{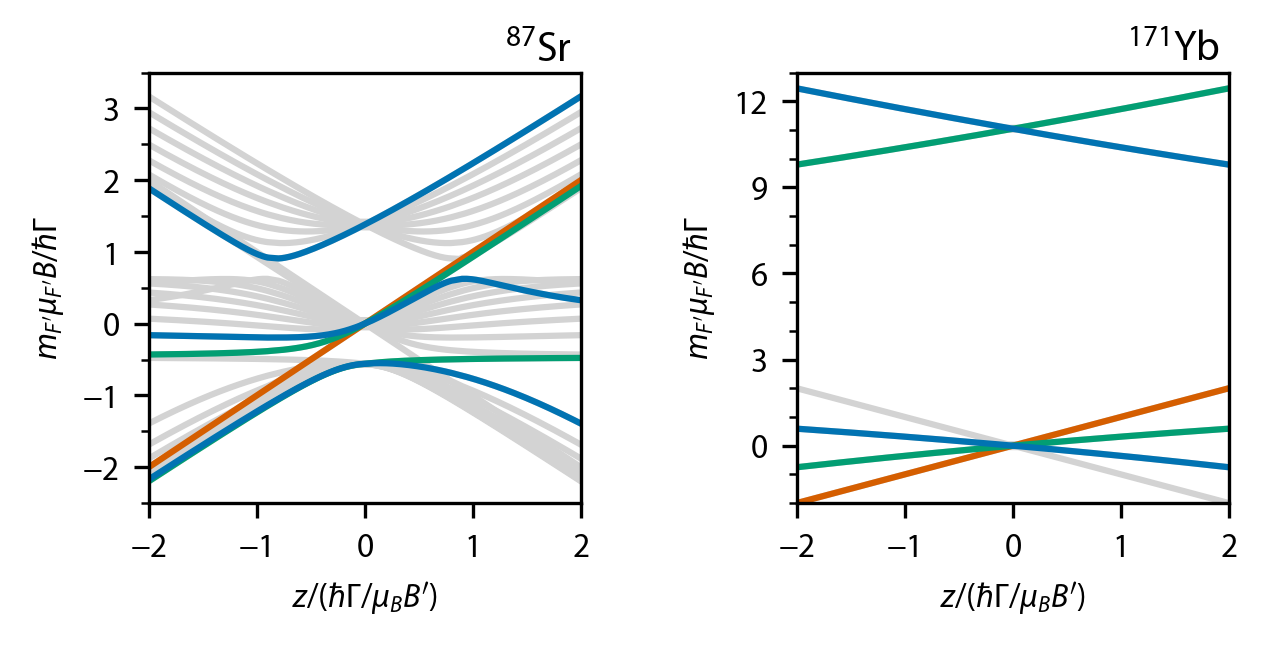}
    \caption{\label{fig:zeemanshift} Zeeman shifts in fermionic alkaline-earth gMOTs.
    Left: Gray curves show the Zeeman shift of \(^{1}P_1\) hyperfine states of \(^{87}\)Sr as a function of axial position \(z\) at \(x=y=0\).
    Right: Gray curves show the Zeeman shift of \(^{1}P_1\) hyperfine states of \(^{171}\)Yb as a function of axial position \(z\).
    In both subplots, the \{red, green, blue\} curves highlight states that are coupled to the ground \(|F, m_F=F\rangle\) state by the \{\(\sigma^+\), \(\pi\), \(\sigma^-\)\} component of the diffracted laser beams.
    }
\end{figure}

The difficulty in observing a \(^{87}\)Sr gMOT is instead caused by a displacement between the bosonic and fermionic gMOT axial equilibrium positions.
Axial confinement in a \(^{1}S_0\,\rightarrow\,^{1}P_1\) gMOT is provided primarily by the \(\sigma^{+}\)-polarized input beam and the \(\pi\)-polarized components of the diffracted beams (see figure~\ref{fig:intro}).
In alkali and bosonic alkaline-earth atoms, transitions driven by \(\pi\)-polarized light exhibit little or no Zeeman shift.
By contrast, \(\pi\)-polarized transitions in fermionic alkaline-earths can exhibit strong Zeeman shifts due to the mismatch in the ground and excited state magnetic moments (see Sec.~\ref{sec:F_to_Fp1}).
Fermionic and bosonic alkaline-earth \(^{1}S_0\,\rightarrow\,^{1}P_1\) gMOTs therefore exhibit significantly different axial restoring forces.
The differing axial restoring forces also shift the axial equilibrium position of a fermionic gMOT away from the bosonic gMOT axial equilibrium position, an effect not seen in conventional \(6\)-beam MOT geometries.
Figure~\ref{fig:trapshift} shows the difference in the axial equilibrium position \(z_{{\rm eq}, A}\) for bosonic alkaline-earth gMOTs (\(A=88,174\)) and fermionic alkaline-earth gMOTs (\(A=87,171\)).
The distance between the axial equilibrium positions of the bosonic and fermionic MOTs often exceeds the typical radius of the trapped atom cloud~\cite{Sitaram2020}.
For \(^{171}\)Yb, the gMOT spin polarizes into the \(|F=1/2, m_F=1/2\rangle\) state and the \(\pi\)-polarized \(|F=1/2, m_F=1/2\rangle\rightarrow |F'=1/2, m_{F'}=1/2\rangle\) transition is Zeeman shifted out of resonance with the diffracted beams as an atom approaches the grating chip, as shown in figure~\ref{fig:zeemanshift}.
The axial equilibrium position of a \(^{171}\)Yb gMOT is therefore shifted away from the quadrupole field zero (toward the chip) compared to a \(^{174}\)Yb gMOT.
For \(^{87}\)Sr, the unresolved excited state hyperfine structure means that MOTs operate in the intermediate field regime of the Zeeman effect~\cite{Kluge1974}.
The \(^{87}\)Sr gMOT still spin polarizes toward the \(|F=9/2, m_F=9/2\rangle\) state, but the \(\pi\)-polarized transition corresponding to \(|F=9/2, m_F=9/2\rangle\rightarrow |F'=9/2, m_{F'}=9/2\rangle\) exhibits minimal Zeeman shift as an atom approaches the grating chip and Zeeman shifts into resonance as an atom moves away from the grating chip (see figure~\ref{fig:zeemanshift}).
Because the \(F'=9/2\) manifold is closer to resonance with the laser beams than the \(F'=11/2\) manifold, a \(^{87}\)Sr gMOT will form closer to the quadrupole field zero than a bosonic strontium gMOT for our trapping parameters.

\begin{figure}
    \center
    \includegraphics[width=\linewidth]{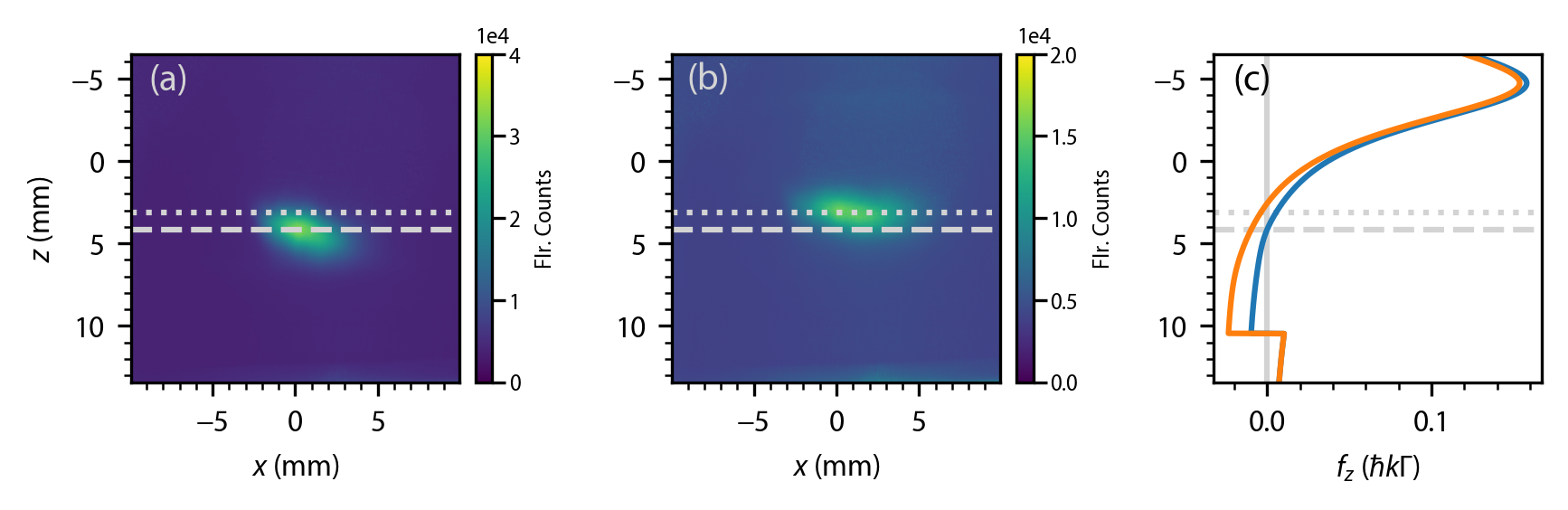}
    \caption{\label{fig:measuredshift} Comparison of measured and computed MOT axial equilibrium position shift for \(^{88}\)Sr and \(^{87}\)Sr.
    Average fluorescence images of \(^{88}\)Sr and \(^{87}\)Sr gMOTs are shown in (a) and (b), respectively.
    The blue (orange) curve in (c) is the axial force profile for a \(^{88}\)Sr (\(^{87}\)Sr) gMOT at \(x=y=0\) and \(v=0\).
    The vertical gray line in (c) guides the eye to the point of zero force in each profile.
    In all subplots, dashed (dotted) horizontal lines denote the fitted axial equilibrium position of the \(^{88}\)Sr (\(^{87}\)Sr) gMOT.
    }
\end{figure}

In our apparatus, the number of atoms confined by the gMOT is sufficiently sensitive to the chip offset \(z_{\text{chip}}\) that an \(^{87}\)Sr MOT is undetectable when the \(^{88}\)Sr gMOT is optimized.
To allow observation of a fermionic strontium gMOT, we replaced the permanent magnets on the apparatus described in Ref.~\cite{Sitaram2020} with a pair of rail-mounted electromagnets that can be repeatably re-positioned with respect to the grating chip.
Changing the position of the electromagnets allows us to adjust \(z_{\text{chip}}\) over a range of several millimeters, which was sufficient for us to observe and optimize fluorescence from a \(^{87}\)Sr gMOT.
Figure~\ref{fig:measuredshift}(a) and figure~\ref{fig:measuredshift}(b) show the average fluorescence image of \(^{88}\)Sr and \(^{87}\)Sr gMOTs, respectively, at \(z_{\text{chip}}\approx 13~\si{\milli\meter}\).
The images in figure~\ref{fig:measuredshift} were acquired with \(s_0\approx 1\), \(\Delta/\Gamma\approx -1.0\), \(w_0\approx 1.2~\si{\centi\meter}\), \(N_b=3\), \(\eta=0.32\), \(\theta_d=27.0(5)\si{\degree}\), and \(B'=5~\si{\milli\tesla\per\centi\meter}\).
The experimentally measured distance between the bosonic and fermionic gMOT equilibrium positions is \(z_{{\rm eq},88}-z_{{\rm eq},87}=1.05(4)~\si{\milli\meter}\), indicating that the effect seen in our simulations exists in the experiment.
(Here, and throughout the paper, parenthetical quantities represent the standard error).
We calculated the axial force profiles for both the \(^{88}\)Sr and the \(^{87}\)Sr gMOT under the experimental conditions, shown in figure~\ref{fig:measuredshift}(c).
The calculations include the additional polarization impurity due to the Stokes parameters of the diffracted laser beams, as measured on a test bench~\cite{Sitaram2020}.
The force profiles predict \(z_{{\rm eq},88}-z_{{\rm eq},87}\approx 1.6~\si{\milli\meter}\).
Small variations in the programmed input and diffracted beam polarization purity can drastically change the expected \(z_{{\rm eq},88}-z_{{\rm eq},87}\).
For example, rotating the azimuthal angle of the diffracted polarizations through \(\pm\pi\) on the Poincar\'{e} sphere produces \(0.3~\si{\milli\meter}\lesssim z_{{\rm eq},88}-z_{{\rm eq},87} \lesssim 1.7~\si{\milli\meter}\).
Because we do not have quantitative \textit{in-situ} measurements of the input beam and diffracted beam polarizations, we believe that the theory is in reasonable agreement with our measurements.

\subsection{Mitigation strategies: effect of diffraction efficiency, diffraction angle, and intensity}
\label{sec:mitigation}

\begin{figure}
    \centering
    \includegraphics{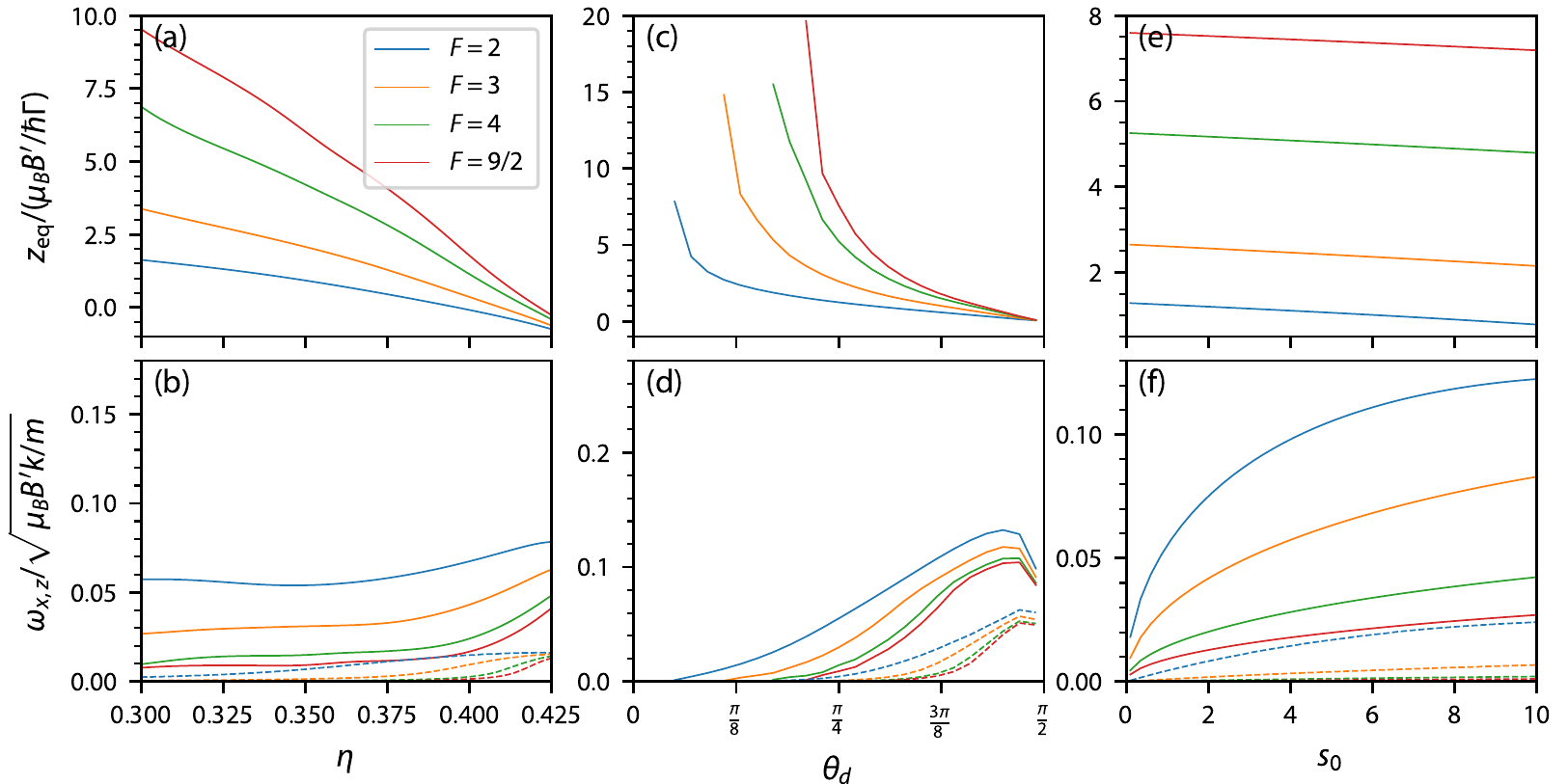}
    \caption{ 
    Effects of diffraction efficiency, diffraction angle, and intensity in type-I gMOTs with \(g_F=1/F\). (a) Axial equilibrium position $z_{\rm eq}$ and (b) trapping frequency along the $z$ direction (solid) and the $x$ direction (dashed curves) of a gMOT vs. diffraction efficiency $\eta$.
    The gMOT has infinite plane-wave beams with $s_0 = 1$, $\Delta/\Gamma=-3/2$, $\theta_d=\pi/4$, $N_b = 3$, $g_{F'}=2/F'$, and $g_F=1/F$ for various $F$.
    The ideal $\eta$ for balanced molasses is $\eta=1/N_b\approx 0.33$. 
    (c) Axial equilibrium position and (d) trapping frequencies vs. diffraction angle $\theta_d$ for the same gMOT parameters as (a) and (b), except $\eta=0.33$.
    (e) Axial equilibrium position and (f) trapping frequencies vs. input beam saturation parameter $s_0$ for the same gMOT parameters as (a) -- (d), with $\eta=0.33$ and $\theta_d=\pi/4$.
    }
    \label{fig:F_to_Fp1_alkali}
\end{figure}

We now consider more thoroughly the feasibility of gMOTs for atoms with level structures that do not satisfy Eq.~\ref{eq:stabMOT}.
The analysis of Sec.~\ref{sec:F_to_Fp1} suggests that such atoms may not be trappable in a gMOT, yet we demonstrated in Sec.~\ref{sec:sryb} that \(^{87}\)Sr, which violates Eq.~\ref{eq:stabMOT}, can be confined by a gMOT because of the unresolved hyperfine structure of its \(^{1}P_1\) state.
In this subsection, we tune the diffraction efficiency \(\eta\), diffraction angle \(\theta_d\), and input beam saturation parameter \(s_0\) beyond the values used in Sec.~\ref{sec:F_to_Fp1} to find stable gMOT operating conditions for atoms that do not satisfy Eq.~\ref{eq:stabMOT} and that have resolved excited state hyperfine structure.

We begin with the alkalis, which have \(g_F=1/F\) and \(g_{F'}=2/F'\) for \(|^{2}S_{1/2},F=I+1/2 \rangle \rightarrow\, |^{2}P_{3/2}, F'=F+1\rangle\) transitions.
Figure~\ref{fig:F_to_Fp1_alkali} shows the axial equilibrium position \(z_{\rm eq}\), axial trapping frequency \(\omega_z\), and transverse trapping frequency \(\omega_x\) for alkali atoms with \(F=2\) (\textit{e.g.} \(^{87}\)Rb), \(F=3\) (\(^{85}\)Rb), \(F=4\) (\(^{133}\)Cs), and \(F=9/2\) (\(^{40}\)K).
We immediately see that combinations of \(\eta\), \(\theta_d\), and \(s_0\) exist for which \(^{133}\)Cs and \(^{40}\)K gMOTs are stable.
As expected given Sec.~\ref{sec:F_to_Fp1}, the confinement of a gMOT, as measured by the trapping frequencies, decreases with \(F\) due to the increasingly adverse effects of spin polarization.

There are several potential ways to increase the trapping forces for large $F$ alkali gMOTs.
One possible way is to increase the power in the diffracted beams by increasing \(\eta\).
As $\eta$ increases, the axial equilibrium position $z_{\rm eq}$ moves closer to the quadrupole field zero because there is more scattering from off-resonant transitions when $z>0$ (see figure~\ref{fig:F_to_Fp1_alkali}(a)).
However, the reduction in \(z_{\rm eq}\) does not increase \(\omega_z\) until \(z_{\rm eq}\approx \hbar\Gamma/\mu_B B'\), as shown in figure~\ref{fig:F_to_Fp1_alkali}(b).

The transverse trapping frequency \(\omega_x\) is smaller than $\omega_z$ but has similar behavior as a function of \(\eta\).
For simplicity, we consider motion along \(x\), where the off-axis Zeeman shift is \(|B(x,0,z_{\rm eq})|=B'\sqrt{z_{\rm eq}^2+x^2/4}\).
For $|x|\gtrsim z_{\rm eq}$, the Zeeman shift is linear in $x$ and the restoring force resembles a textbook one-dimensional MOT~\cite{Foot2005}.
For $|x|\lesssim z_{\rm eq}$, the lack of an appreciable Zeeman shift produces little confining force or---depending on the values of $\theta_d$, $\eta$ and $s_0$---a small anti-confining force.
In figure~\ref{fig:F_to_Fp1_alkali}, we calculate an ``average'' $\omega_x$ by taking the difference ${\rm max}(f_x)-{\rm min}(f_x)$ divided by the distance between the two, which is approximately $4 z_{\rm eq}$ for $z_{\rm eq}\gg\hbar\Delta/\mu_B B'$ and approximately $4 \hbar \Delta/\mu_B B'$ for  $z_{\rm eq}\ll \hbar\Delta/\mu_B B'$.
This choice masks an important complication: realizing a large \(F\), alkali gMOT requires that the patterned grating area, input beam size, magnetic field gradient, and \(z_{\rm chip}\), combine such that the distance from $\mathbf{r} = (0,0,z_{\rm eq})$ to the edge of the overlap region of the beams is larger than roughly $2 z_{\rm eq}$ to ensure transverse confinement.

A more promising approach to increased confining forces in large $F$ alkali gMOTs is tuning the diffraction angle $\theta_d$, which preserves balanced molasses for subsequent sub-Doppler cooling.
The resulting axial equilibrium positions and trapping frequencies are shown in figure~\ref{fig:F_to_Fp1_alkali}(c) and figure~\ref{fig:F_to_Fp1_alkali}(d), respectively.
As $\theta_d$ gets larger, the projection of the diffracted beam polarization on to the $z$ axis gains $\pi$ and $\sigma^-$ character at the expense of $\sigma^+$, helping to reduce spin polarization.
As a result, the $\sigma^-$ component has a larger effect when it becomes resonant at $z>0$, leading to a deeper negative force peak at $z>0$ that pushes $z_{\rm eq}$ back toward zero.
This occurs despite the fact that momentum kicks along $z$ due to the diffracted beams are reduced as \(\theta_d\) gets larger.
We expect that gMOTs of \(^{133}\)Cs and \(^{40}\)K may be attainable at \(\theta_d\gtrsim 3\pi/8\) over a wide range of \(\eta\), though the large gratings and beam sizes needed to maintain trapping volume could prove impractical.

Finally, the trapping forces in a gMOT can be improved by increasing the input beam saturation parameter \(s_0\).
Both \(\omega_z\) and \(\omega_x\) roughly double when the input saturation raises from \(s_0=1\) to \(s_0>5\) (see figure~\ref{fig:F_to_Fp1_alkali}(f)), as transitions due to the \(\pi\) and \(\sigma^-\) components of the diffracted beams enter the strongly saturated regime.
Boosting \(s_0\) has a minimal effect on \(z_{\rm eq}\), as shown in figure~\ref{fig:F_to_Fp1_alkali}(e).
We expect that experiments attempting to produce gMOTs for alkalis that violate Eq~\ref{eq:stabMOT}, should use large \(s_0\), large \(\theta_d\), and \(\eta\approx 1/N_b\) to preserve balanced molasses.

\begin{figure}
    \centering
    \includegraphics{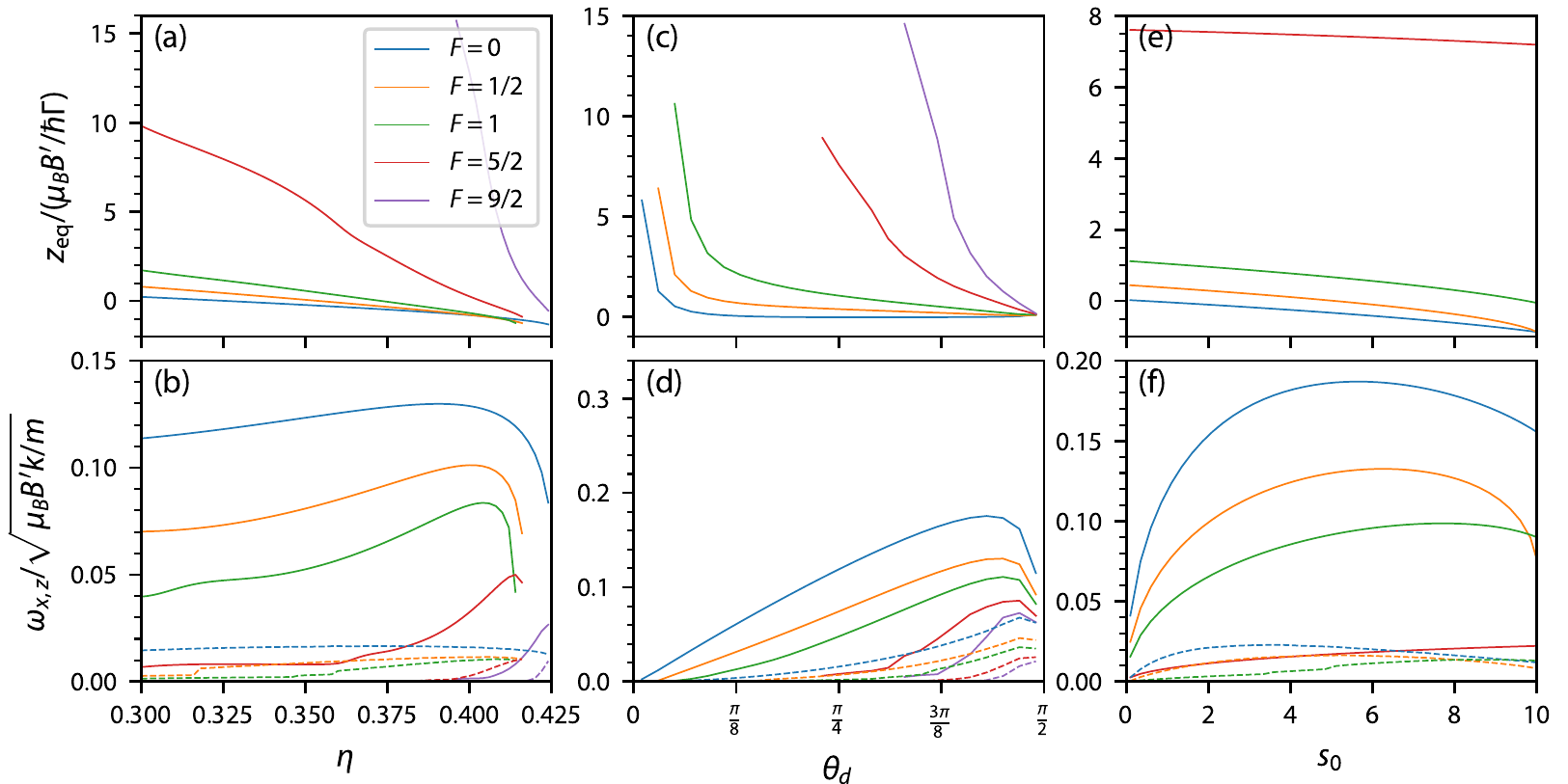}
    \caption{
    Effects of diffraction efficiency, diffraction angle, and intensity in type-I gMOTs with \(g_F=0\). (a) Axial equilibrium position $z_{\rm eq}$ and (b) trapping frequency along the $z$ direction (solid) and the $x$ direction (dashed curves) of a gMOT vs. diffraction efficiency $\eta$.
    The gMOT has infinite plane-wave beams with $s_0 = 1$, $\Delta/\Gamma=-3/2$, $\theta_d=\pi/4$, $N_b = 3$, $g_{F'}=1/F'$, and $g_F=0$ for various $F$.
    The ideal $\eta$ for balanced molasses is $\eta=1/N_b\approx 0.33$. 
    (c) Axial equilibrium position and (d) trapping frequencies vs. diffraction angle $\theta_d$ for the same gMOT parameters, except $\eta=0.33$.
    (e) Axial equilibrium position and (f) trapping frequencies vs. input beam saturation parameter $s_0$ for the same gMOT parameters as (a) -- (d), with $\eta=0.33$ and $\theta_d=\pi/4$.
    }
    \label{fig:F_to_Fp1_eta}
\end{figure}

We now turn our attention to gMOTs operating on transitions with $g_F\approx 0$ and resolved excited state hyperfine structure, which arise for alkaline-earth elements.
As with an alkali gMOT, one way to improve confinement is to increase the power in the diffracted beams by increasing \(\eta\).
The improved confinement is clear in figure~\ref{fig:F_to_Fp1_eta}(a) and figure~\ref{fig:F_to_Fp1_eta}(b).
%, which show both the equilibrium position $z_{\rm eq}$ and trapping frequency in the $z$ direction $\omega_z$ vs. diffraction efficiency $\eta$ for a variety of different $F$.
Again, as $\eta$ increases, the axial equilibrium position $z_{\rm eq}$ moves closer to the origin.
The axial trapping frequency $\omega_z$ is roughly constant with \(\eta\) for \(z_{\rm eq} > \hbar\Gamma/\mu_B B'\) and increases slowly with \(\eta\) for \(z_{\rm eq} < \hbar\Gamma/\mu_B B'\).
For large values of $\eta$, $\omega_z$ maximizes and subsequently drops rapidly as the force from the diffracted beams overwhelms the force from the input beam, making the MOT unstable.
As in figure~\ref{fig:F_to_Fp1_alkali}, we calculate an ``average'' $\omega_x$ by taking the difference ${\rm max}(f_x)-{\rm min}(f_x)$ divided by the distance between the two.
Around \(z_{\rm eq}=0\), \(\omega_x\) has a broad peak as a function of \(\eta\).
%Once again, the behavior of \(\omega_x\) falls into three regimes as a function of \(\eta\).}
%For intermediate \(\eta\), \(\omega_x\) becomes negative due to the small rotation of the large magnetic field at \(z=z_{\rm eq}\) and confining forces are only substantial for \(x\gtrsim z_{\rm eq}\).
%For small \(\eta\), \(\omega_x\) becomes positive again because anti-trapping \(\sigma^+\) transitions from the diffracted beams are suppressed by the higher intensity input beam.}
Increasing confinement via higher diffraction efficiency is quite promising; the only drawback is it unbalances the optical molasses that might be used in any subsequent sub-Doppler cooling~\cite{Lee2013}.

Rather than compromising the balanced molasses for sub-Doppler cooling, one can tune the diffraction angle $\theta_d$.
The resulting axial equilibrium positions and trapping frequencies are shown in figure~\ref{fig:F_to_Fp1_eta}(c) and figure~\ref{fig:F_to_Fp1_eta}(d).
Again, as $\theta_d$ gets larger, the projection of the diffracted beam polarization on to the $z$ axis gains $\pi$ and $\sigma^-$ character, which leads to reduced spin polarization that pushes $z_{\rm eq}$ toward zero.
At $\eta=0.33$, large diffraction angles, $\theta_d\gtrsim3\pi/8$, are required  to trap $F=5/2$ ($^{173}$Yb) and $F=9/2$ ($^{87}$Sr) atoms.
The large grating and beams needed to maintain a reasonable capture volume likely make increasing $\theta_d$ alone impractical.
However, increasing both $\eta$ and $\theta_d$ appears ideal, in combination, a stable \(^{1}S_0\,\rightarrow\,^{3}P_1\) transition gMOT may be attainable for $^{173}$Yb and $^{87}$Sr. 

% \begin{figure}
%     \center
%     \includegraphics{F2_to_F3_trapping_frequencies_vs_pol}
%     \caption{\label{fig:F2_to_F3_freq} Trapping frequencies for a $F=5/2\rightarrow F'=7/2$ gMOT over the Poincar\'{e} sphere of the diffracted beams for $s_0 = 1$, $\Delta/\Gamma=-3/2$, $\theta_d=\pi/4$, $N_b = 3$, $\eta=0.32$, $g_F=0$, and $g_{F'}=1/F'$.}
% \end{figure}

The trapping forces in a \(g_F\approx0\) gMOT can also be improved by tuning \(s_0\).
The dependence of \(\omega_z\) and \(\omega_x\) on \(s_0\) shown in figure~\ref{fig:F_to_Fp1_eta}(f) differs from the alkali case shown in figure~\ref{fig:F_to_Fp1_alkali}(f).
Both \(\omega_z\) and \(\omega_x\) peak between \(1<s_0<10\).
%The transverse trapping frequency peaks near \(z_{\rm eq}=0\) (see figure~\ref{fig:F_to_Fp1_eta}(e)), where the dependence of the Zeeman shift on \(x\) is maximized.
The maximum \(\omega_z\) occurs when \(\pi\) and \(\sigma^-\) transitions from the diffracted beams become saturated strongly enough to begin dragging the entire force profile below zero.
%For large \(F\), \(z_{\rm eq}\) varies weakly with \(s_0\) and the behavior of the trapping frequencies is similar to the alkali case.
For \(F=5/2\), \(\omega_x\approx0\) for all \(s_0\) in figure~\ref{fig:F_to_Fp1_eta}(f).
We also note that \(F=9/2\) does not appear in figure~\ref{fig:F_to_Fp1_eta}(e) or figure~\ref{fig:F_to_Fp1_eta}(f) because the \(F=9/2\) gMOT is not stable for any \(s_0\) at \(\eta=0.33\) and \(\theta_d=\pi/4\).

The \(F=1/2\) gMOT simulations in figure~\ref{fig:F_to_Fp1_eta}(d) indicate that a \(^{171}\)Yb gMOT should be stable, contradicting Sec.~\ref{sec:sryb}.
However, the simulations of Sec.~\ref{sec:sryb} used laser beams with Gaussian spatial modes, while the simulations in figure~\ref{fig:F_to_Fp1_eta} used laser beams with uniform intensity.
When a gMOT employs Gaussian beams, the intensity of the diffracted beams that push toward (away from) the gMOT equilibrium position decreases (increases) as an atom moves off the \(z\) axis.
When \(F=0\), the transverse intensity variation of Gaussian diffracted beams only reduces the transverse restoring force~\cite{McGilligan2015}.
For \(F>0\), the spin polarization combines with the intensity variation to promote anti-trapping.
%The gMOT is spin polarized toward \(|m_F=F\rangle\), which has larger Clebsch-Gordan coefficients for \(\sigma^+\) transitions (driven by beams pushing out of the gMOT) than for \(\sigma^-\) transitions (driven by beams pushing into the gMOT); the transverse intensity variation promotes anti-trapping.
For gMOTs operating at low \(s_0\), as is typical for \(^1S_0\rightarrow\, ^1P_1\) transition alkaline-earth MOTs~\cite{Maruyama2003,Xu2003a}, the anti-confining forces due to the Gaussian beam shape can be large enough to destabilize the gMOT.
Transverse confinement can be restored by significantly increasing the radius of the input laser beam.
We anticipate that power efficient operation of a \(^{171}\)Yb \(^1S_0\rightarrow\, ^1P_1\) transition gMOT may require uniform illumination of the grating chip, as has recently been demonstrated for \(^{87}\)Rb and \(^{88}\)Sr in Refs.~\cite{McGehee2021, Bondza2022}.

\begin{figure}
    \center
    \includegraphics{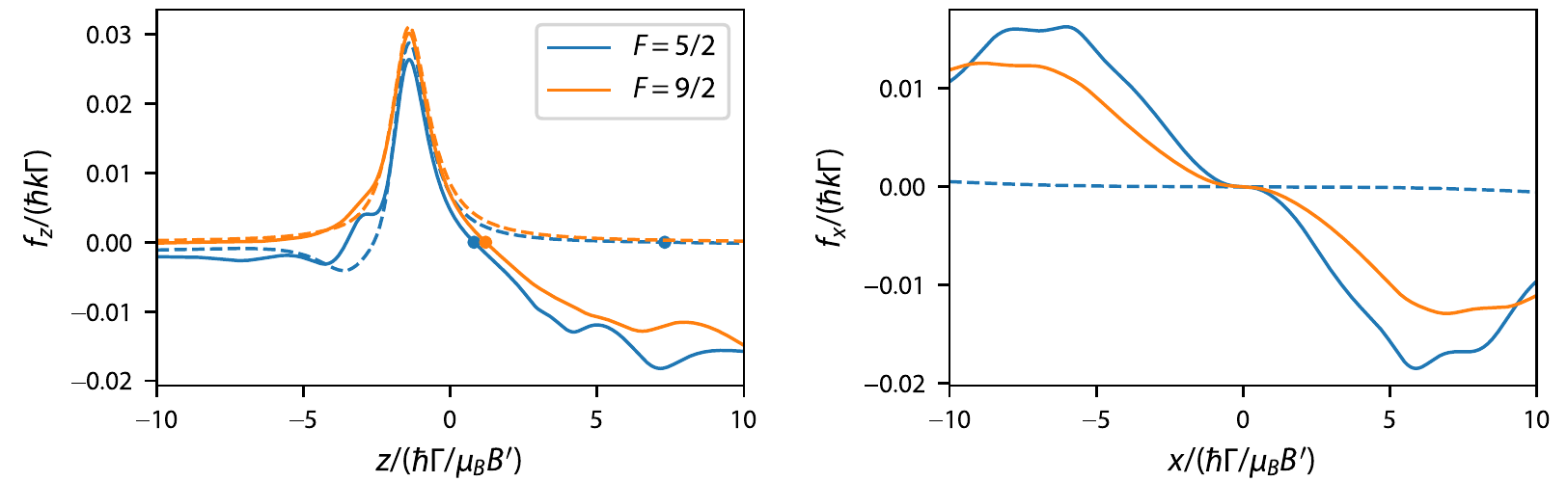}
    \caption{\label{fig:twocolor_gf0}
    Forces in single-frequency and dual-frequency gMOTs with \(g_F=0\). 
    Left: force $f_z(0,0,z)$ along $z$ at $v=0$ for one-color gMOT (dashed) formed on a $F\rightarrow F'=F+1$ transition with $s_0 = 1$, $\Delta/\Gamma=-3/2$, $\theta_d=\pi/4$, $N_b = 3$, $g_{F'}=1/F'$ and a dual-frequency MOT (solid) with the same parameters, but with an additional, oppositely polarized input beam with $\Delta_b/\Gamma=+3$.
    Points show the stable equilibrium point $f_z(0,0,z_{\rm eq})=0$.
    Right: force $f_x(x,0,z_{\rm eq})$ along $x$ at $v=0$ for the same gMOTs as on the left.}
\end{figure}

A $g_F\approx 0$ gMOT could also be stabilized by introducing a second input laser beam at a different optical frequency.
Following the conventional six-beam stirring approach for \(^{87}\)Sr, the second beam could be red-detuned from a resolved \(F\rightarrow F'=F\) transition (see figure~\ref{fig:zeemanshift}).
However, for large $F$, we do not expect conventional stirring to prevent spin polarization because the \(F\rightarrow F'=F\) input beam still biases the population toward the \(|m_F=F\rangle\) Zeeman state and, once an atom is pumped into that state, all transitions from the stirring diffracted beams will be far from resonance~\cite{Mukaiyama2003}.
There are two unconventional ways that a second input beam could stabilize a $g_F\approx 0$ gMOT at large \(F\).
First, the second beam could exploit the spin polarization and drive a \(F\rightarrow F'=F-1\) transition.
Because \(F\rightarrow F'=F-1\) transitions in alkaline-earth fermions have negative \(g_{F'}\), the \(\sigma^-\) component of the \(F\rightarrow F'=F-1\) diffracted beams will tune into resonance with the $|m_F=F\rangle$ state at $z>0$, pulling \(f_z\) below zero and stabilizing the gMOT.
Second, we can borrow the ``dual-frequency'' MOT technique\footnote{In a ``dual-frequency'' MOT at least one transition is addressed by both a red-detuned and a blue-detuned laser beam.} from molecule laser cooling by blue-detuning a second \(\epsilon^-\)-polarized input beam by \(\Delta_b\) from a \(F\rightarrow F'=F\) or \(F\rightarrow F'=F+1\) transition~\cite{Tarbutt2015} (and see Sec.~\ref{sec:typeII}).
In the dual-frequency configuration, the blue-detuned diffracted beams shift into resonance with the $|m_F=F\rangle$ state at $z>0$, once again pulling \(f_z\) below zero and stabilizing the gMOT.
The forces in both single-frequency and dual-frequency \(g_F=0\) gMOTs, when the blue-detuned beam drives \(F\rightarrow F'=F+1\), are shown in figure~\ref{fig:twocolor_gf0} for \(F=5/2\) (\(^{173}\)Yb) and \(F=9/2\) (\(^{87}\)Sr).
We note that alkaline-earth \(^{1}S_0\,\rightarrow\,^{3}P_1\) MOTs are typically frequency modulated at frequencies comparable to \(\Gamma\) and therefore operate out of equilibrium.
Optical Bloch equation simulations, including frequency modulation, will be necessary to confirm that a second input beam can stabilize \(^{1}S_0\,\rightarrow\,^{3}P_1\) gMOTs.
Such simulations are beyond the scope of this work and will be the subject of a future publication.

\section{Type-II MOTs}
\label{sec:typeII}
The on-axis spin polarizing effect seen in type-I MOTs is amplified in type-II MOTs. 
In particular, atoms in type-II gMOTs tend to be pumped into a dark state that does not vary dramatically with position along the axis of the MOT.
This nearly-spatially independent dark state tends to eliminate any potential restoring force on axis in the simplest type-II gMOT configurations.

\subsection{Simple angular momentum cases}
\label{sec:type-II_simple}

\subsubsection{\texorpdfstring{$F\rightarrow F'=F$}{F to F'=F} transitions}
\label{sec:F_to_F}

\begin{figure}
  \center
  \includegraphics{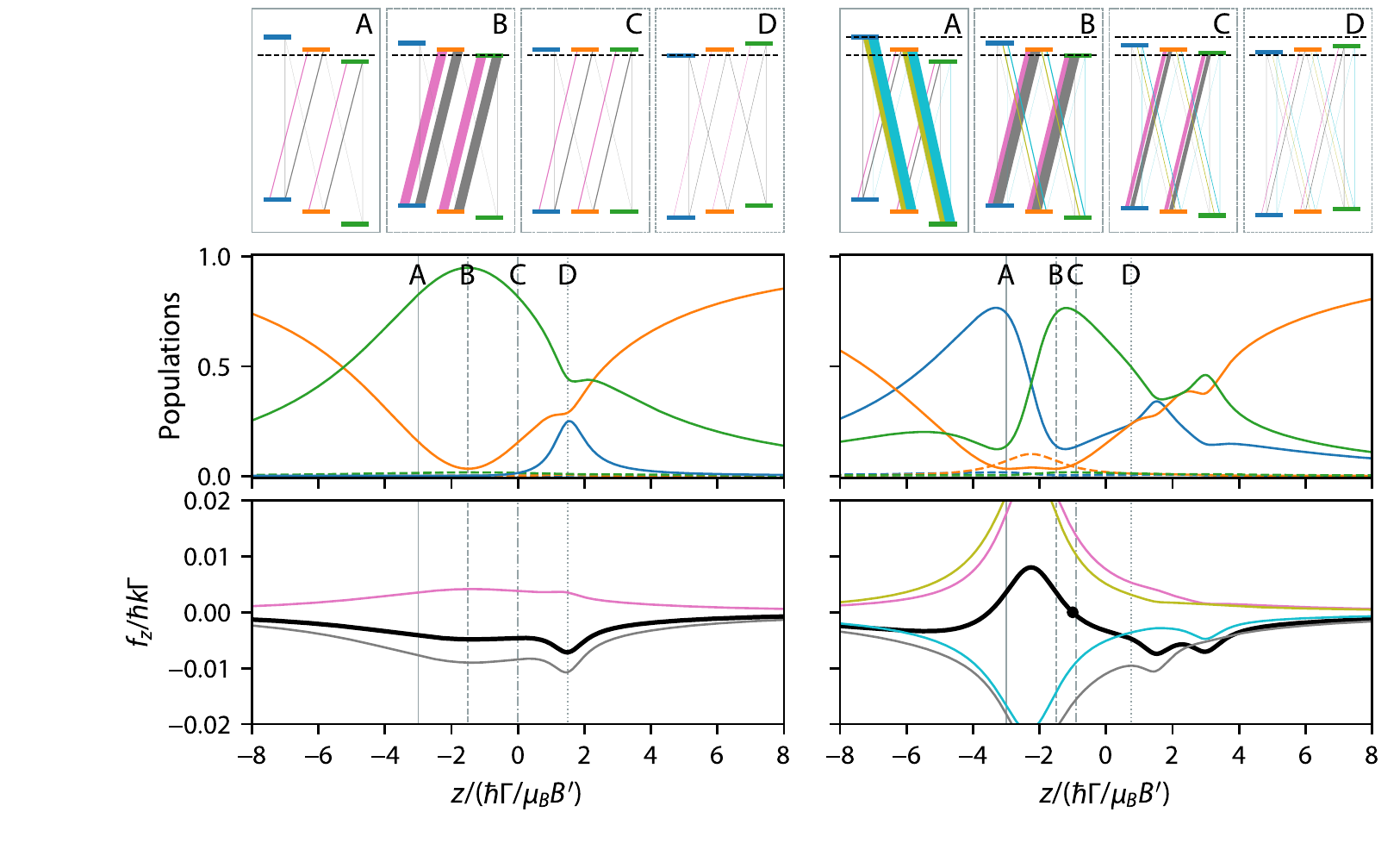}
  \caption{\label{fig:F_to_F_level} Optical pumping (top row), populations (middle row) and forces (bottom row) along the axial direction for a standard gMOT (left column) and dual-frequency gMOT (right column) operating on a $F=1\rightarrow F'=1$ transition.
  In all subplots, $x=y=0$ and $v=0$.
  The MOT parameters are $\Delta/\Gamma=-3/2$, $s_0 = 1$, $N_b = 3$, $\eta=0.33$, $\theta_d=\pi/4$, $g_F=1$, and $g_{F'}=1$.
  The second, blue-detuned beam has the same intensity and detuning $\Delta_b/\Gamma = +3$.
  The labels A, B, C, D indicate the positions at which the level diagrams and optical pumping strengths in the top row are calculated.
  The thickness of the magenta, gray, olive, and cyan lines denotes pumping rates for the red-detuned incident, red-detuned diffracted, blue-detuned incident, and blue-detuned diffracted laser beams, respectively.
  The populations of the ground (solid) and excited (dashed) states in the middle row are colored to match the $m_F$ color coding of the level diagrams.
  Total force in the bottom row plots is the black curve, which is a sum of the force from the incident red-detuned beam (magenta), the diffracted red-detuned beams (gray), the incident blue-detuned beam (olive), and the diffracted blue-detuned beams (cyan).
  Black dots on the total force curves show the location where the MOT will form ($f_z = 0$).}
\end{figure}

The force for a gMOT operating on an $F=1 \rightarrow F'=1$ transition, where the diffracted beams have opposite polarization as the incident beam, is shown in the left column of figure~\ref{fig:F_to_F_level}.
The ground state Land\'e $g$-factors are $g_F=g_{F'}=1$, a case that is partially reflective of some ${\rm X}^2\Sigma \rightarrow {\rm B}^2\Sigma$ transitions in alkaline-earth-monofluoride molecules. 
The magnetic-field-independent $\pi$ transitions tend to pump the atoms toward the $|m_F=0\rangle$ dark state, most easily observed at large $|z|$, where $N_{m_F=0}\rightarrow 1$ and $f_z\rightarrow 0$.
At $z=-\hbar \Delta/\mu_B B'$, the dark state of $|m_F=0\rangle$ is perturbed by the $\sigma^-$ component of the diffracted beams, pumping atoms into $|m_F=-1\rangle$ and exerting a small, negative $f_z$. 
At $z=\hbar \Delta/\mu_B B'$, the dominant $\sigma^+$ light becomes resonant and changes the dark state to $|m_F=1\rangle$.
Once in $|m_F=1\rangle$, an atom needs to scatter a $\pi$ or $\sigma^-$ photon from the diffracted beams to be excited out of the dark state.
In the former case, the atom moves to $|m_{F'}=1\rangle$.
From $|m_{F'}=1\rangle$, the atom can decay to the dark $|m_F=1\rangle$ state or bright $|m_F=0\rangle$ state with roughly equal probability.  
If it decays into $|m_F=0\rangle$, it will most likely be pumped back by the resonant $\sigma^+$ light to $|m_{F'}=+1\rangle$ receiving a momentum kick from either the incoming or the reflecting beam.
On average for $\eta=1/N_b$ and $\theta_d=\pi/4$, the atom receives more force from the diffracted beams through this process and therefore $f_z < 0$.
Because $f_z < 0$ for all $z$, no MOT will form.

Following the approach of Sec.~\ref{sec:mitigation}, we find two paths to creating the positive $f_z$ required for a MOT.
The first path is increasing the diffraction angle to reduce the $\hat{z}$ momentum kick from the diffracted beams and simultaneously strengthen their $\pi$-polarized component.
Achieving a stable MOT in this way requires very large diffraction angles, $\theta_d\gtrsim 80^\circ$ for $F=1\rightarrow F'=1$.
For $F>1$, the necessary $\theta_d$ increases further, making diffraction angle engineering alone even more impractical than in Sec.~\ref{sec:mitigation}.
%Second, adjusting the polarization of the diffracted light can decrease the $\sigma^+$ component relative to $\pi$ component in the diffracted beams.
%As in Sec.~\ref{sec:mitigation}, achieving a stable MOT by manipulating the diffracted polarization likely requires metasurface optics.

The second, and more promising, path to a type-II gMOT is the dual-frequency MOT~\cite{Tarbutt2015a}.
As previously noted in Sec.~\ref{sec:mitigation}, a dual-frequency MOT adds a second, blue-detuned laser with opposite circular polarization to repump Zeeman dark states.
We take the detuning of the second beam to be $\Delta_b/\Gamma = 3$, which yields a reasonable compromise between enhanced trapping force and reduced damping force.
The operation of an $F=1\rightarrow F'=1$ dual-frequency gMOT is shown in the right panel of figure~\ref{fig:F_to_F_level}.
In molecular six-beam MOTs, dual-frequency operation sacrifices damping force to increase spatial confinement.
In a gMOT, dual-frequency operation completely salvages the situation, rapidly changing the dark state between the locations of resonance for the blue and red detuned beams, generating a large peak with $f_z>0$ and a $z_{\rm eq}<0$.

\begin{figure}
  \center
  \includegraphics{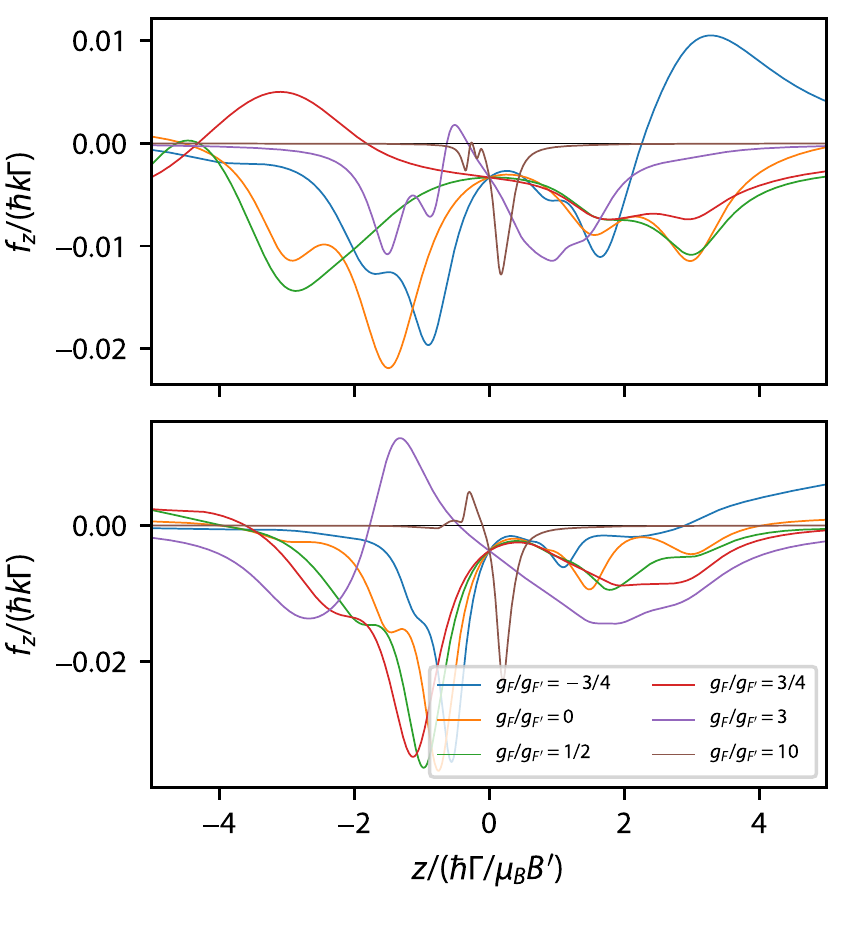}
  \caption{\label{fig:F_to_F_gfactors} Axial force $f_z$ as a function of position $z$ in a $F=1\rightarrow F'=1$ (top) and $F=2\rightarrow F'=2$ (bottom) dual-frequency gMOT with $\Delta/\Gamma = -3/2$, $s_0=1$, $N_b=3$, $\eta=0.33$, $\theta_d=\pi/4$, and $g_{F'} = 1/F'$.
  The blue-detuned beam has the same intensity as the red-detuned beam and detuning $\Delta_b/\Gamma = +3$.
  In both subplots, $x=y=0$ and $v=0$.
  }
\end{figure}

We explore the robustness of dual-frequency gMOTs for several relevant ratios of $g_F$ to $g_{F'}$ in figure~\ref{fig:F_to_F_gfactors}.
Alkali atoms with $I=3/2$ (\textit{e.g.}, $^7$Li, $^{23}$Na, and $^{87}$Rb) with a type-II MOT operating on the $D_2$ line have either $g_F/g_{F'}=\pm3/4$.
Likewise, alkali atoms with $I=3/2$ have $g_F/g_{F'}=3$ for a type-II MOT operating on the $D_1$ line.
Alkaline-earth atoms like Sr, Ca, and Yb generally have $g_F/g_{F'}\approx 0$. 
Alkaline-earth-monofluoride MOTs operating on ${\rm X}\rightarrow {\rm A}$ transitions typically have $g_F/g_{F'}\gg 1$.
A hypothetical MOT operating on a ${\rm X}\rightarrow {\rm B}$ transition in CaF would have $g_F/g_{F'}\approx 0.8$ for the upper $F=1\rightarrow F'=1$ transition and $g_F/g_{F'}\approx -0.3$ for the lower $F=1\rightarrow F'=1$ transition~\cite{Tarbutt2015a}.
The dual-frequency gMOT operation only produces a axial confinement when $g_F/g_{F'}=3/4$ or $g_F/g_{F'}=3$ (for $F=1\rightarrow F'=1$ transitions) and $g_F/g_{F'}=3$ or $g_F/g_{F'}=10$ (for $F=2\rightarrow F'=2$ transitions).
The dual-frequency gMOT may also be stable for for $F=1\rightarrow F'=1$ transitions when $g_F/g_{F'}=-3/4$ after flipping the input laser beam polarization to $\epsilon^-$.

\subsubsection{\texorpdfstring{$F\rightarrow F'=F-1$}{F to F'=F-1} transitions}
\label{sec:F_to_Fm1}

\begin{figure}
  \center
  \includegraphics{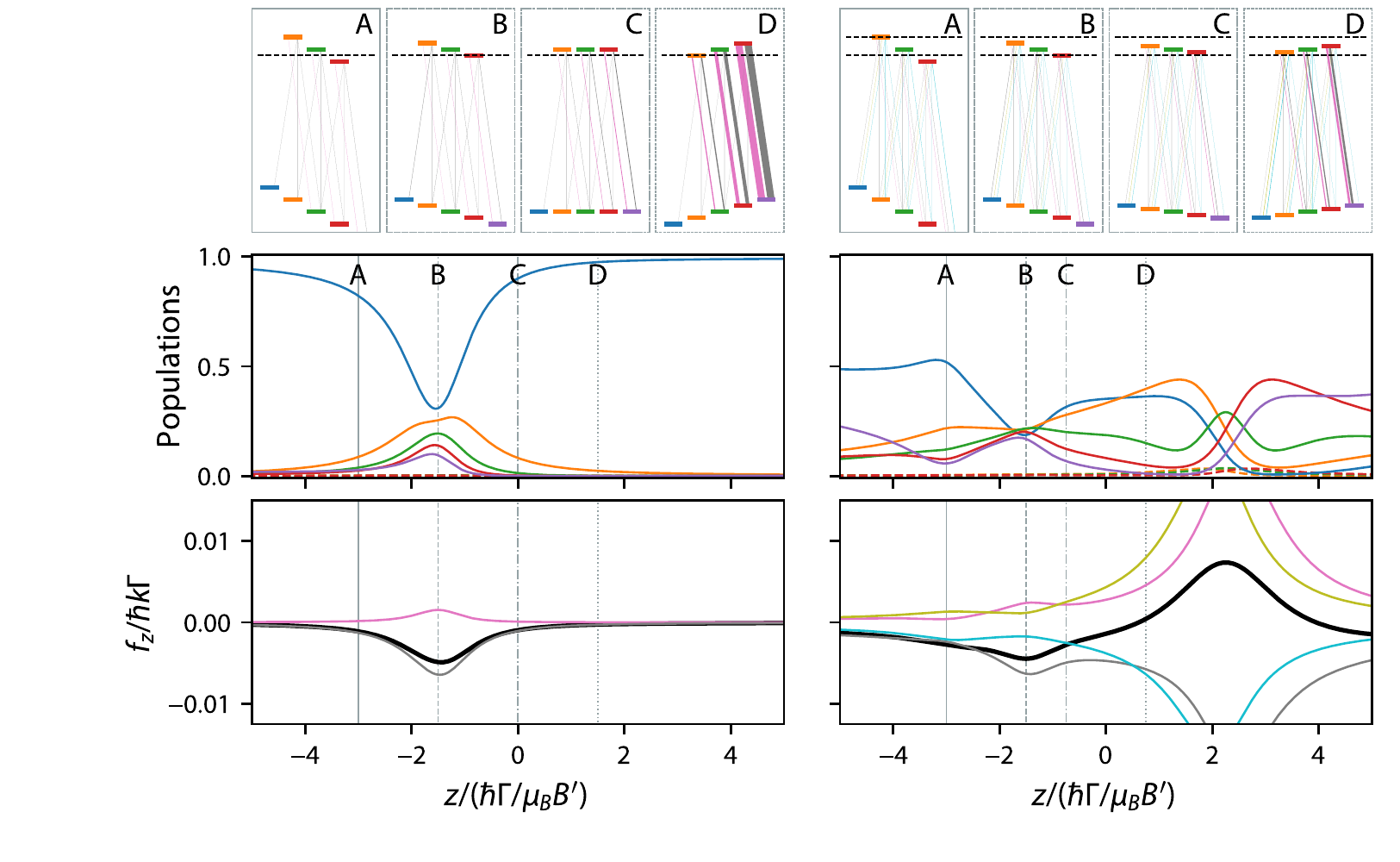}
  \caption{\label{fig:F_to_Fm1_level} Optical pumping (top row), populations (middle row) and forces (bottom row) along the axial direction in the $F=2\rightarrow F'=1$ grating for a standard gMOT (left column) and dual-frequency gMOT (right column).
  In all subplots, $x=y=0$ and $v=0$.
  The MOT parameters are $s_0 = 1$, $\Delta/\Gamma=-3/2$, $\theta_d=\pi/3$, $N_b = 3$, $g_F=1$, and $g_{F'}=1$.
  The second, blue-detuned beam has the same intensity and detuning $\Delta_b/\Gamma = +3$.
  The labels A, B, C, D indicate the positions at which the level diagrams and optical pumping strengths are calculated.
  Colorings are the same as in figure~\ref{fig:F_to_F_level}.
  %Black dots on the total force curves show the location where the MOT will form ($f_z = 0$).
  }
\end{figure}

The spin polarization into a single $|m_F\rangle$ state creates a rather dire situation for $F \rightarrow F'=F-1$ transitions, precluding the formation of a stable MOT.
Figure~\ref{fig:F_to_Fm1_level} shows the level diagrams, populations, and axial forces for a $F=2 \rightarrow F'=1$ transition with $g_F/g_{F'}=1$ in a gMOT with $\epsilon^-$ incident light (switched from $\epsilon^+$ according to rules of Ref.~\cite{Tarbutt2015}).
The now dominant $\epsilon^-$ light in both the incident and diffracted beams drive $\sigma^-$ transitions, biasing the population toward $|m_F=-2\rangle$.
The population only depolarizes when the now suppressed $\sigma^+$ component of the diffracted beams comes into resonance at $z = \hbar \Delta/\mu_B B'$.
At this location, the $\sigma^+$ transitions driven by the diffracted beams cause scattering from the $|m_F=-2\rangle$ state lead to $f_z<0$.
Because there is no location where scattering from the incident beam is larger than that of the diffracted beam, no stable MOT will form.

\begin{figure}
  \center
  \includegraphics{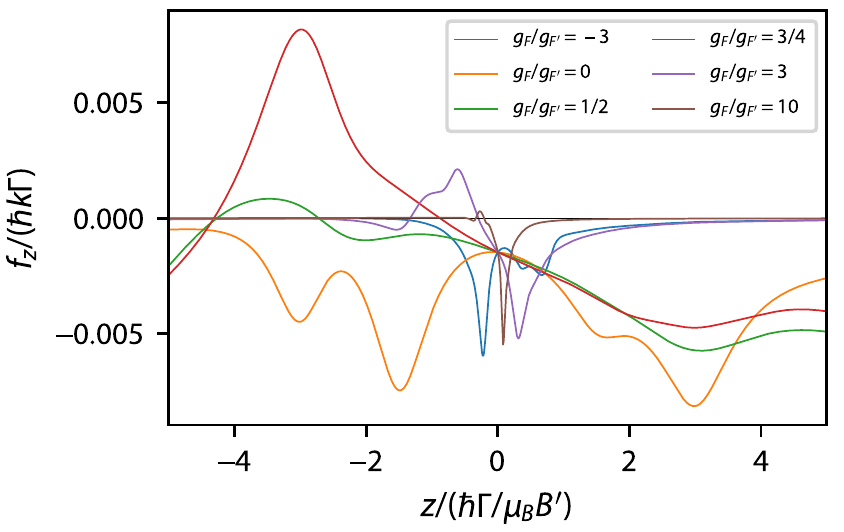}
  \caption{\label{fig:F_to_Fm1_gfactors} Axial force $f_z(0,0,z)$ as a function of position $z$ at $v=0$ in a $F=2\rightarrow F'=1$ dual-frequency gMOT with $\Delta/\Gamma = -3/2$, $s_0=1$, $N_b=3$, $\eta=0.33$, $\theta_d=\pi/4$, $g_{F'} = 1$, and various $g_F/g_{F'}$.}
\end{figure}

Dual-frequency operation again salvages the type-II gMOT.
The additional blue-detuned beam drives population out of the dark state as evidenced by the more complicated dependence of $N_{m_F}$ with $z$ in the right column of figure~\ref{fig:F_to_Fm1_level}.
Curiously, stable MOT operation requires the opposite incident polarization as predicted by Ref.~\cite{Tarbutt2015}, namely $\epsilon^+$.
Reversing the incident polarization would mirror the total force in the right column of figure~\ref{fig:F_to_Fm1_level} about $z=0$, and produce a stable trapping point at $z_{\rm eq} < 0$.
Figure~\ref{fig:F_to_Fm1_gfactors} shows the performance of a dual-frequency, $F=2\rightarrow F'=1$ gMOT for several relevant ratios of $g_F$ to $g_{F'}$.
Alkali atoms with $I=3/2$ have $g_F/g_{F'}=3/4$ in a $F=2\rightarrow F'=1$ MOT operating on the $D_2$ line and they have $g_F/g_{F'}=-3$ in a $F=2\rightarrow F'=1$ MOT operating on the $D_1$ line.
As in the $F=1\rightarrow F'=1$ gMOT considered earlier, alkaline-earth atoms have $g_F/g_{F'}\approx 0$, alkaline-earth-monofluorides have $g_F/g_{F'}\gg 1$ for ${\rm X}\rightarrow {\rm A}$ transitions, and alkaline-earth-monofluorides have $g_F/g_{F'}\approx 0.5$ for ${\rm X}\rightarrow {\rm B}$ transitions. %($g_F/g_{F'}\approx 0.5$ for CaF).
Dual-frequency gMOT operation only produces a stable $F=2\rightarrow F'=1$ MOT when $g_F/g_{F'}=3/4$ or $g_F/g_{F'}=3$.

\subsection{Experimental cases: \texorpdfstring{$D_1$}{D1} MOTs in \texorpdfstring{$^7$Li}{7Li}}
\label{sec:exptype2}

The results shown in Sec.~\ref{sec:F_to_F} and~\ref{sec:F_to_Fm1}, suggest that gMOTs utilizing type-II, or even hybrid type-I and type-II, transitions will only work for a very specific set of level structures.
Consider gMOTs with Li atoms on the $D_1$ ($^2\mbox{S}_{1/2}\rightarrow\, ^2\mbox{P}_{1/2}$) transition.
There are three possible type-II transitions: (1) $F=1\rightarrow F'=1$ with $g_F/g_{F'}=3$, (2) $F=2\rightarrow F'=2$ with $g_F/g_{F'}=3$, and (3) $F=2\rightarrow F'=1$ with $g_F/g_{F'}=-3$.
Of the three type-II transitions, the most promising for a gMOT is case (2) according to figure~\ref{fig:F_to_F_gfactors} and figure~\ref{fig:F_to_Fm1_gfactors}.
Case (1) appears marginal at best and case (3) looks impossible.

Despite the rather grim prospects of success, we nonetheless attempted to make type-II, and hybrid type-I and type-II, gMOTs using transitions on the $D_1$ line of $^7$Li.
The experimental apparatus and grating chip are described in detail in Refs.~\cite{Barker2019, Barker2022}.
To summarize, the grating chip has a \(1.1~\si{\centi\meter}\) radius patterned area, which produces $N_b=3$ beams with a diffraction angle of $\theta_d \approx 42^\circ$ and diffraction efficiency $\eta \approx 0.37$.
With a normally incident, left-hand circularly polarized input beam, our grating produces near perfect circular polarization on reflection: the Stokes parameter corresponding to circular polarization is $V=0.92(1)$.
%Given the parameters of our grating and the results of Sec.~\ref{sec:F_to_F}, it would be surprising if any MOT were to form on this transition.

\begin{figure}
    \center
    \includegraphics{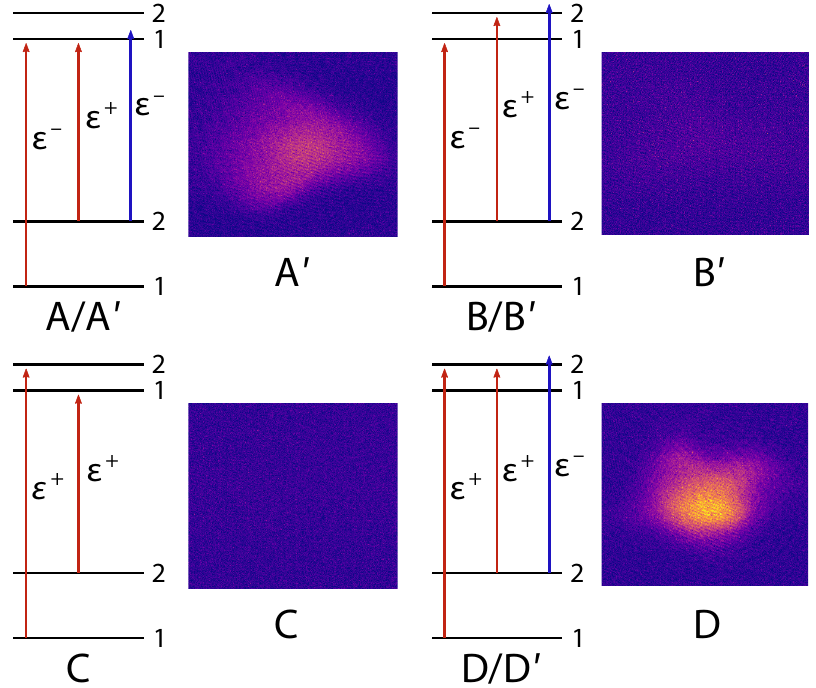}
    \caption{\label{fig:D1_galleria}
    Hyperfine level diagrams and absorption images of the four types of $D_1$-line MOTs.
    Red and blue arrows denote red-detuned and blue-detuned input laser beams, respectively.
    When a blue-detuned beam is present, we use prime notation to indicate dual-frequency gMOT operation as described in the text.
    The absorption images were taken after 10~ms in the MOT at magnetic field gradients of \(1.9\)~mT/cm (type-A$'$), \(6.0\)~mT/cm (type-B$'$), \(2.3\)~mT/cm (type-C), and \(2.3\)~mT/cm (type-D).
    We have also tested the type-D$'$ MOT; an absorption image is not shown.
    }
\end{figure}

There are two hyperfine states ($F'=1,2$) in the excited $D_1$ manifold of $^7$Li (since $I=3/2$).
The ground state $g$-factors are $g_{F=1}= -1/2$ and $g_{F=2}=1/2$; likewise, the excited state $g$-factors are $g_{F'=1}=-1/6$ and $g_{F'=2}=1/6$.
Given the total angular momenta of the ground and excited states, there are four possible combinations of frequencies and polarizations that can create MOTs on the $D_1$ transition.
Ref.~\cite{Flemming1997} explored these four combinations in a six-beam MOT geometry, and found that all could form a MOT.
The level diagrams of these four combinations are shown in figure~\ref{fig:D1_galleria}.
When we use only red-detuned light (red arrows in figure~\ref{fig:D1_galleria}), we label the four combinations as types A--D.
A blue-detuned beam (blue arrows in figure~\ref{fig:D1_galleria}) can be added to any of the combinations to make a dual-frequency MOT, which we denote with a prime, \textit{e.g.}, type-A$'$.

Our type-II gMOTs are loaded from a standard type-I gMOT formed on the Li $D_2$ transition ($^2\mbox{S}_{1/2}\rightarrow\, ^2\mbox{P}_{3/2}$).
Our laser system allows us to test $D_1$ MOT types A$'$, B$'$, C, D, and D$'$.
Further details on the $D_1$ laser system and MOT loading procedure contained in \ref{sec:apparatus}.

We attempted to make a type-B$'$ gMOT using an incident beam with three frequency components.
One frequency component drives the $F=1\rightarrow F'=1$ transition with saturation parameter $s_0\approx 0.9$, detuning $\Delta/\Gamma = -1.0$, and $\epsilon^-$ polarization.
The other two frequency components drive the $F=2\rightarrow F'=2$ transition: one component has $\Delta/\Gamma = -2.0$, $s_0\approx 1.8$, and polarization $\epsilon^+$; the other component has $\Delta/\Gamma = 3.0$, $s_0\approx 0.2$, and polarization $\epsilon^-$.
The magnetic field gradient gradient was 6.0~mT/cm.
No gMOT was observed (see figure~\ref{fig:D1_galleria}).

We attempted to make a type-C MOT using an incident beam with two frequency components with the same $\epsilon^+$ polarization.
Both frequency components have detuning $\Delta/\Gamma=-1.0$.
The saturation parameters are $s_0\approx 1.3$ for the  $F=1\rightarrow F'=2$ frequency component and $s_0 \approx 1.0$ for the $F=2\rightarrow F'=1$ frequency component.
The magnetic field gradient gradient was scanned from 1.5~mT/cm to 6.0~mT/cm.
No gMOT was observed (see figure~\ref{fig:D1_galleria}).

We did observe a type-A$'$ gMOT, as shown in figure~\ref{fig:D1_galleria}.
The type-A$'$ gMOT uses an input beam with three frequency components.
One component drives the $F=1\rightarrow F'=1$ transition with saturation parameter $s_0\approx 1.1$, detuning $\Delta/\Gamma = -1.5$, and $\epsilon^-$ polarization.
The other two frequency components drive the $F=2\rightarrow F'=1$ transitions: One component has $\Delta/\Gamma = -1.5$, $s_0\approx 1.8$, and $\epsilon^+$ polarization; the other has $\Delta/\Gamma = 3.0$, $s_0\approx 0.2$, and $\epsilon^-$ polarization.
The image in figure~\ref{fig:D1_galleria} shows a type-A$'$ MOT at $B'=1.9$~mT/cm.
Type-A$'$ gMOTs are observed at all $B'$ between 1.5~mT/cm and 6~mT/cm, but the lifetime decreases with larger $B'$.

We also successfully observed a hybrid type-I/type-II (type-D) gMOT.
The type-D gMOT uses an input beam with two frequency components, each with $\Delta/\Gamma = -1.5$.
The frequency component addressing the type-I $F=1\rightarrow F'=2$ transition has $s_0\approx 1.1$; the component addressing the type-II $F=2\rightarrow F'=2$ transition has $s_0\approx 1.4$.
The image in figure~\ref{fig:D1_galleria} shows a type-D gMOT at $B'=2.3$~mT/cm.
We see Type-D gMOTs at all $B'$ between 1.5~mT/cm and 6~mT/cm, but the lifetime decreases with larger $B'$ (as with type-A$'$).
Adding an additional blue-detuned beam, making a type-D$'$ MOT, reduced the initial atom number in the MOT without increasing the lifetime.
We therefore focus our subsequent analysis solely on the type-D MOT.

The existence of the type-A$'$ and type-D gMOTs is quite surprising, if we na\"ively consider the analysis of Sec.~\ref{sec:type-II_simple}.
For the type-A$'$ gMOT, the $F=1\rightarrow F'=1$ transition should provide no confinement according to figure~\ref{fig:F_to_F_level} and we are driving the $F=2\rightarrow F'=1$ transition with the incorrect polarization according to figure~\ref{fig:F_to_Fm1_level}.
For the type-D gMOT, figure~\ref{fig:F_to_F_level} suggests that there should be no confinement; only a type-D$'$ gMOT would be expected to yield confinement on both $F=1\rightarrow F'=2$ and $F=2\rightarrow F'=2$ transitions (see figure~\ref{fig:F_to_F_gfactors}).
Both initial analyses of the $D_1$-line gMOTs assume that forces due to transitions from the $F=1$ and $F=2$ ground states are additive, which, given the presence of dark states, is incorrect.

\begin{figure}
    \center
    \includegraphics{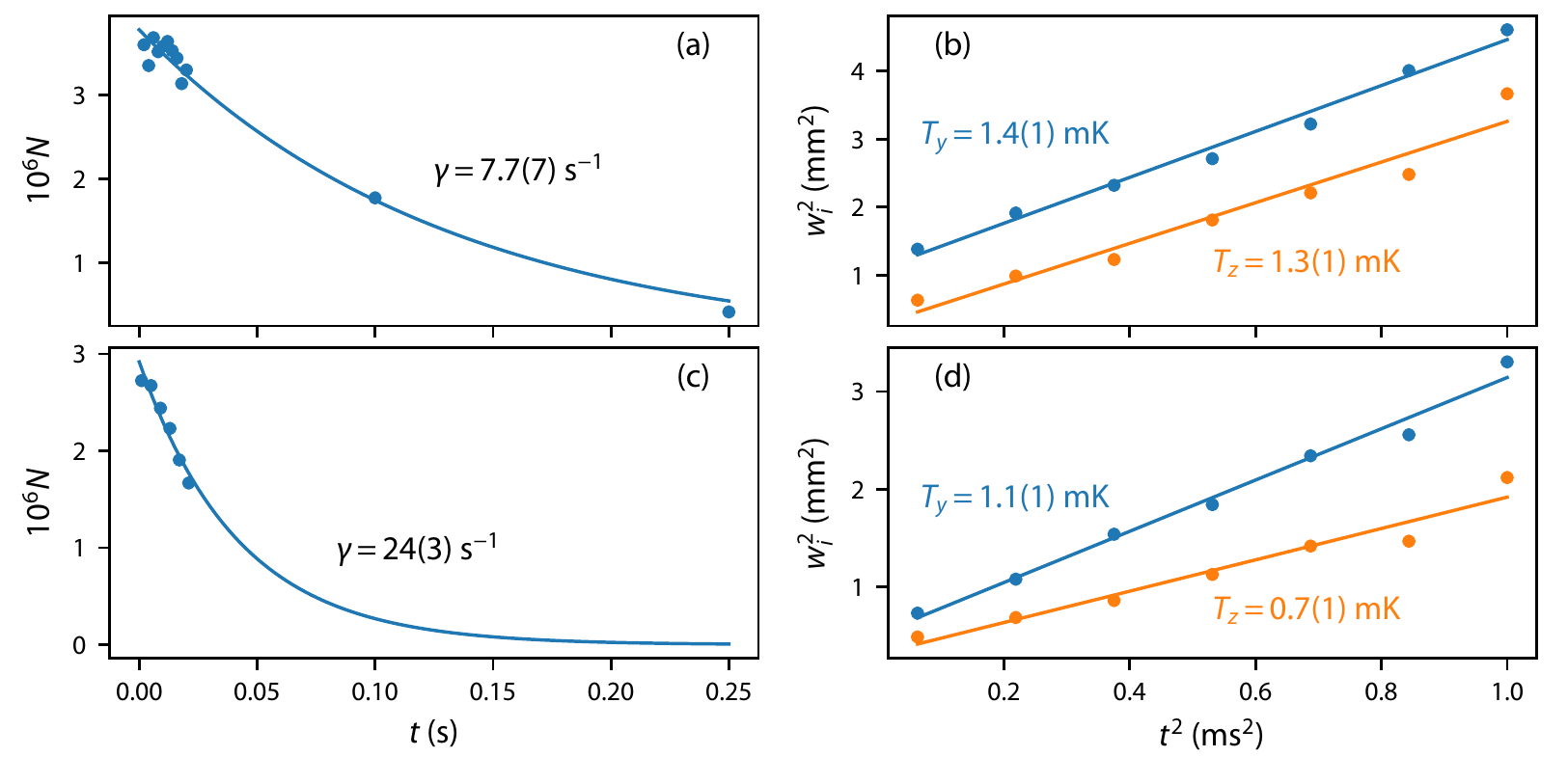}
    \caption{\label{fig:Li_D1_properties} Properties of the observed type-II Li gMOTs.  (a) Atom number vs. time with exponential fit to extract the lifetime for type-A$'$ gMOT.  (b) Width squared $w_{i=y,z}^2$ vs. time squared $t^2$ with a linear fit to extract the temperature in both directions for the type-A$'$ gMOT. (c) and (d) are the same as (a) and (b), but for the type-D gMOT.}
\end{figure}

To characterize their curious existence, we measured the atom loss rates and temperatures of the $D_1$-line gMOTs.
The type-A$'$ and type-D gMOTs have single-body loss rates of $\gamma = 7.7(7)$~s$^{-1}$ and $\gamma = 24(3)$~s$^{-1}$, as shown in figure~\ref{fig:Li_D1_properties}(a) and (c).
The measured loss rates are larger than the vacuum-limited loss rate of the type-I gMOT, which is approximately 1.3~s$^{-1}$.
Rather than decaying to a non-zero equilibrium atom number, both type-II gMOTs have no atoms remaining in the trap after 1~s.
This indicates that they have no appreciable capture velocity -- if atoms were being continuously loaded into the trap from the atomic source, the equilibrium number $N$ would be $N=R/\gamma$, where, in this context, $R$ is the loading rate and $\gamma$ the loss rate from the trap.
The negligible capture velocity of type-II gMOTs is particularly striking in the context of the six-beam counterpart~\cite{Flemming1997}, which was loaded directly from a hot atomic vapor.
The temperatures of the gMOTs are shown in figure~\ref{fig:Li_D1_properties}(b) and (d).
Both type-II gMOTs are hotter than the type-I gMOT.
The temperature along the transverse direction $T_y$ is hottest, with temperatures of $T_y/T_D\approx 10$ and $T_y/T_D\approx 8$, for the type-A$'$ and type-D MOTs, respectively ($T_D = \hbar\Gamma/2k_B$ is the Doppler temperature).
The higher temperatures compared to a type-I MOT are consistent with observations from six-beam type-II MOTs. 

\begin{figure}
    \center
    \includegraphics{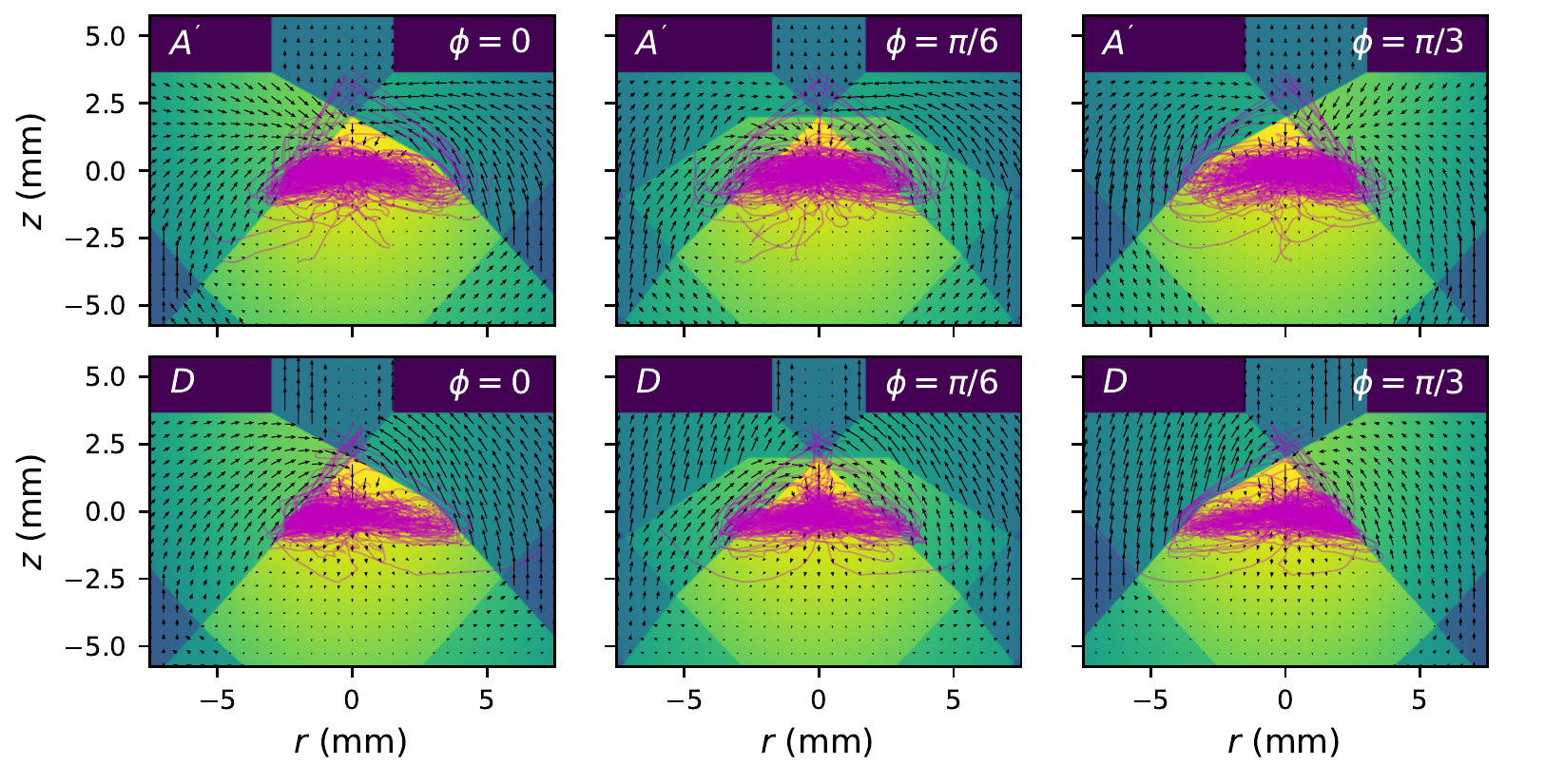}
    \caption{\label{fig:Li_D1_sims} Simulation of the type-A$'$ (top row) and type-D $^7$Li gMOTs.
    The columns correspond to the same data viewed along three different cut-through planes, oriented along the azimuthal angle $\phi$.
    Each panel shows the total intensity of lasers, both incident and diffracted, in a colormap from blue to yellow.
    The vector field encodes the direction and magnitude of the equilibrium forces exerted on  a stationary atom.
    The magenta curves show the simulated trajectories of atoms in the MOT.
    The corresponding view angle for figure~\ref{fig:D1_galleria} is $\phi=\pi/6$.}
\end{figure}

To complete our understanding of the type-A$'$ and type-D gMOTs, we simulated them using our rate equation model, including the full level structure of the $D_1$ line.
We use the same incident beam size, relative detunings and intensities as the experiment.
We use $B'=2$~mT/cm both type-A$'$ and type-D simulations since the experimental trap loss rate varies slowly with $B'$.
In the simulations, we include the finite size of the grating, the Gaussian nature of the incident beams, and reflection of the beams from the grating given the incident Gaussian beam.
We neglect diffraction effects in the reflection, which will have the effect of softening the edges of the diffracted beams as they propagate away from the grating.
The resulting laser intensity map $s_{\rm total}(\mathbf{r})$ is shown as the color plots in figure~\ref{fig:Li_D1_sims}.
In the center, there is the a region of maximum intensity, which corresponds to the overlap region of the incident and three diffracted beams.
At the top, there is a portion of the incident beam that transmits through the hole in the chip~\cite{Barker2019}.
Other regions correspond to the overlap of the incident beam with either one or two of the diffracted beams.

We first calculate the equilibrium forces for a stationary atom, shown as the vector field in figure~\ref{fig:Li_D1_sims}.
As we expect from the single-color gMOT results in Sec.~\ref{sec:type-II_simple}, the forces in the overlap region tend to point away from the grating chip, toward $-z$ (which is parallel to gravity).
The blue-detuned frequency component of the type-A$'$ gMOT does not produce a restoring force, perhaps due to its low $s_0$.
%This is also true for the type-A$'$ gMOT, where the blue-detuned component is only one of on the two transitions and has a relatively low $s_0$.
We then place 128 atoms at the origin with all eight hyperfine ground states equally populated and evolve the atoms according to Eq.~\ref{eq:populations}-\ref{eq:force}.
To include the stochastic force due to spontaneous emission, random recoils of magnitude $2 \hbar k$ are periodically inserted with a probability $N_e \Gamma \Delta t$, where $N_e$ is the total population in the excited state and $\Delta t$ is the variable timestep in the simulation.
To ensure that scattering events are not ``missed'', we constrain $\Delta t$ such that $\Delta t < 0.1/(N_e\Gamma)$ (\textit{i.e.} the probability of a random recoil in any timestep is always less than $0.1$).
Each atom's motion is simulated for 10.4~ms, unless the atoms' position $\mathbf{r}$ exceeds twice the length of a parallel vector $\mathbf{r}'$ that extends from the magnetic field zero to the edge of the overlap volume.
Such an atom is considered ``lost''.

The simulated atom trajectories are shown in figure~\ref{fig:Li_D1_sims} as magenta lines.
The simulations show both type-A$'$ and type-D gMOTs forming, despite the equilibrium forces indicating a net downward force in the overlap region.
While the simulated shapes do not agree with those in figure~\ref{fig:D1_galleria}, other properties are in reasonable agreement.
For example, the simulations predict, based on the loss of some atoms, loss rates of 33.5~s$^{-1}$ and 12~s$^{-1}$ for type-A$'$ and type-D gMOTs, respectively. 
They also predict an increase of about a factor of 4 in loss rate as $B'$ is increased from 2~mT/cm to 6~mT/cm, similar to what we observe in the experiment.
Finally, the simulated $T_y=1.2$~mK and $T_y=0.97$~mK for type-A$'$ and type-D gMOTs, respectively, also agree with our measurements.

The simulations indicate that the $D_1$-line gMOTs exist because the grating and incident beam are not infinite.
Specifically, the atoms in the simulation do experience a net force that tends to eject them down and out of the overlap region.
In fact, all atoms, even the ones not lost, leave the overlap region of the beams.
However, once outside of the overlap region, the forces they experience are dramatically different: the forces tend to push the atoms toward positive $z$ and back toward the overlap region.
Thus, the $D_1$-line gMOTs are not acting like stable MOTs: the atoms are constantly leaving and reentering the overlap region.
Instead, $D_1$-line gMOTs act more like atom recyclers, keeping the atoms in looped trajectories.

We also simulated type-C and type-B$'$ gMOTs for completeness.
The simulated type-C MOT does not efficiently recycle the atoms along the loop trajectories; instead, once atoms are in the overlap region, they are ejected upwards.
Our simulations predict that a type-B$'$ atom recycler should form.
It is unclear why we were unable to observe it experimentally.

Finally, we also repeated these simulations using the full optical Bloch equations~\cite{Eckel2022}, to ensure that the forces were not underestimated due to dark states rotating into bright states by Zeeman precession.
Simulations of the full optical Bloch equations show no significant difference compared to those of the rate equations.
Thus, we conclude that Zeeman precession of the ground states does not play a significant role in the formation of these gMOTs.

\section{Discussion}
\label{sec:discussion}

We have studied the operation of gMOTs with real atoms and molecules both experimentally and theoretically, using a rate equation model.
%The asymmetric laser beam geometry leads to spin polarization of the trapped gas.
%For alkalis, the spin polarization does not significantly perturb the trap, and may even be desirable in certain applications.
Our model predicts that high \(F\) atoms and many laser-coolable molecules will be extremely difficult to trap in the asymmetric laser beam geometry of a gMOT.
The challenge in trapping both arises due to spin polarization into states that do not scatter photons from the diffracted or incident beams.
For high \(F\) atoms, gMOTs using gratings with large diffraction angles and high diffraction efficiency should be able to confine isotopes with high nuclear spin (such as \(^{133}\)Cs, \(^{40}\)K, $^{173}$Yb, and $^{87}$Sr).
The rate equations also suggest that dual-frequency operation will stabilize high-\(F\) fermionic alkaline-earth gMOTs, but optical Bloch equation simulations are needed to confirm this prediction.
We believe the more promising approach to realizing fermionic alkaline-earth gMOTs operating on dipole-forbidden transitions is to work with low-nuclear-spin isotopes, such as $^{171}$Yb.
For open shell molecules (such as MgF, CaF, SrF, and YO), dual-frequency gMOT operation appears mandatory, rather than beneficial, as in six-beam molecule MOTs.
Unfortunately, dual-frequency operation only salvages the gMOT for a narrow range of ground and excited state $g$-factors. 
As a result, molecular type-II gMOTs will be highly species dependent and ascertaining their stability will likely require simulations of the full transition level structure on a case-by-case basis.
Closed shell molecules (such as BH, AlF, and AlCl), which can form type-I gMOTs, are more promising, but are subject to the same concerns as alkali and alkaline-earth elements regarding high nuclear spin.

Our results will inform future development of compact quantum devices based on laser-cooled atoms or molecules.
Type-II gMOTs appear to be of little utility, so quantum devices based on open shell molecules will prefer compact MOT geometries that preserve the symmetry of the six-beam MOT, such as photonic-integrated-circuit MOTs or pyramid MOTs~\cite{Lee1996, Isichenko2023, Ropp2023}.
Type-I gMOTs are effective for initial laser cooling and can load ``bright'' or ``gray'' optical  molasses to achieve sub-Doppler temperatures~\cite{Lee2013, Barker2022}.
As a result, type-I gMOTs are promising platform for devices such as quantum memories, atom interferometers~\cite{Lee2022}, or vacuum sensors~\cite{Ehinger2022}. 
In these applications, the spin polarization of the gMOT may even be beneficial.
Because spin polarization destabilizes high-nuclear-spin alkaline-earth gMOTs, optical lattice clocks employing gMOT atom sources should be easiest to realize with bosonic isotopes or $^{171}$Yb.

\section*{Acknowledgements}

We thank E. Altunta\c{s} and W. McGehee for their careful reading of the manuscript.
This work was supported by the National Institute of Standards and Technology.
D.S.B., N.N.K., E.B.N., and S.P.E. have US Patent
11,291,103. D.S.B. and S.P.E. have filed US Provisional Patent
63/338,047.

\section*{Data Availability}

The data that support the findings of this study are available
from the corresponding author upon reasonable request.

\appendix
\section{Lithium type-II gMOT apparatus}
\label{sec:apparatus}

Our type-II gMOTs are loaded from a standard type-I gMOT formed on the Li $D_2$ transition ($^2\mbox{S}_{1/2}\rightarrow\, ^2\mbox{P}_{3/2}$), which is described in detail in Ref.~\cite{Barker2019, Barker2022}.
The type-I MOT is loaded with detuning $\Delta/\Gamma=-5.1$, $s_0\approx 4.9$, and an axial magnetic field gradient of $B' \approx 6$~mT/cm.
After loading, the type-I MOT is ``compressed'': the detuning is rapidly changed to either $\Delta/\Gamma=-2.0$, to load types B and D, or $\Delta/\Gamma=-1.5$, to load types A and C, and the intensity is dropped to  $s_0 \approx 0.5$, causing a rapid density increase and temperature reduction.
The final axial and radial temperatures of the type-I compressed MOT are approximately 400~$\mu$K and 700~$\mu$K, respectively.
Finally, the laser light addressing the $D_2$ transition is shuttered and the $D_1$ light introduced.
At the same time, the magnetic field gradient is also instantaneously changed to a value between 1.5~mT/cm and 6~mT/cm.
The incident beam has a $18$~mm $1/e^2$ radius, apertured down to an 11~mm radius.
Saturation parameters $s_0$ refer to peak saturation parameter in the incident beam.

Light for the $D_1$ type-II MOT comes from a laser locked to either the $F=2\rightarrow F'=1$ transition (type A), $F=2\rightarrow F'=2$ transition (types B and D), or the $F=1\rightarrow F'=2$ transition (type-C).
Light from the laser is passed through two different acousto-optic modulators (AOMs) for independent frequency and intensity control, coupled into independent polarization-maintaining optical fibers, and then recombined on a polarizing beam splitting cube to form the incident beam.
Thus, the light from these two AOMs have opposite circular polarization.
An electro-optic modulator (EOM) installed after one of the AOMs generates light addressing transitions from the $F=1$ state (types A, B, and D) or $F=2$ state (type-C)~\cite{Barker2022}.
The light from the combined EOM/AOM path is always accompanied by an additional optical frequency component with the same polarization that addresses transitions from $F=2$ (types A, B, and D) or from $F=1$ (type-C).
For the type-A and type-B gMOTs, the extra component is incorrectly polarized at red detuning (see figure~\ref{fig:D1_galleria}), so we instead detune it to the blue and make a dual-frequency gMOT.

\section*{References}
\bibliography{main}

\providecommand{\newblock}{}
\begin{thebibliography}{10}
\expandafter\ifx\csname url\endcsname\relax
  \def\url#1{{\tt #1}}\fi
\expandafter\ifx\csname urlprefix\endcsname\relax\def\urlprefix{URL }\fi
\providecommand{\eprint}[2][]{\href{http://arxiv.org/abs/#2}{arXiv:#2}}

\bibitem{Zhang2015}
Zhang J, Tandecki M, Collister R, Aubin S, Behr J~A, Gomez E, Gwinner G, Orozco
  L~A, Pearson M~R and Sprouse G~D 2015
  \href{http://dx.doi.org/10.1103/physrevlett.115.042501}{ {\em Physical Review
  Letters\/} {\bf 115} 042501 }
  \urlprefix\url{https://doi.org/10.1103/physrevlett.115.042501}

\bibitem{Miyake2019}
Miyake H, Pisenti N~C, Elgee P~K, Sitaram A and Campbell G~K 2019
  \href{http://dx.doi.org/10.1103/physrevresearch.1.033113}{ {\em Physical
  Review Research\/} {\bf 1} 033113 }
  \urlprefix\url{https://doi.org/10.1103/physrevresearch.1.033113}

\bibitem{Hinkley2013}
Hinkley N, Sherman J~A, Phillips N~B, Schioppo M, Lemke N~D, Beloy K, Pizzocaro
  M, Oates C~W and Ludlow A~D 2013 {\em Science\/} {\bf 341} 1215
  \urlprefix\url{http://science.sciencemag.org/content/341/6151/1215}

\bibitem{Ohmae2021}
Ohmae N, Takamoto M, Takahashi Y, Kokubun M, Araki K, Hinton A, Ushijima I,
  Muramatsu T, Furumiya T, Sakai Y, Moriya N, Kamiya N, Fujii K, Muramatsu R,
  Shiimado T and Katori H 2021 \href{http://dx.doi.org/10.1002/qute.202100015}{
  {\em Advanced Quantum Technologies\/} {\bf 4} 2100015 }
  \urlprefix\url{https://doi.org/10.1002/qute.202100015}

\bibitem{Cassella2017}
Cassella K, Copenhaver E, Estey B, Feng Y, Lai C and M\"{u}ller H 2017
  \href{http://dx.doi.org/10.1103/physrevlett.118.233201}{ {\em Physical Review
  Letters\/} {\bf 118} 233201 }
  \urlprefix\url{https://doi.org/10.1103/physrevlett.118.233201}

\bibitem{Rudolph2020}
Rudolph J, Wilkason T, Nantel M, Swan H, Holland C~M, Jiang Y, Garber B~E,
  Carman S~P and Hogan J~M 2020
  \href{http://dx.doi.org/10.1103/PhysRevLett.124.083604}{ {\em Phys. Rev.
  Lett.\/} {\bf 124} 083604 }
  \urlprefix\url{https://link.aps.org/doi/10.1103/PhysRevLett.124.083604}

\bibitem{Eckel2018b}
Eckel S, Kumar A, Jacobson T, Spielman I~B and Campbell G~K 2018
  \href{http://dx.doi.org/10.1103/PhysRevX.8.021021}{ {\em Phys. Rev. X\/} {\bf
  8} 021021 }
  \urlprefix\url{https://link.aps.org/doi/10.1103/PhysRevX.8.021021}

\bibitem{Subhankar2019}
Subhankar S, Wang Y, Tsui T~C, Rolston S~L and Porto J~V 2019
  \href{http://dx.doi.org/10.1103/PhysRevX.9.021002}{ {\em Phys. Rev. X\/} {\bf
  9} 021002 }
  \urlprefix\url{https://link.aps.org/doi/10.1103/PhysRevX.9.021002}

\bibitem{Graham2022}
Graham T~M, Song Y, Scott J, Poole C, Phuttitarn L, Jooya K, Eichler P, Jiang
  X, Marra A, Grinkemeyer B, Kwon M, Ebert M, Cherek J, Lichtman M~T, Gillette
  M, Gilbert J, Bowman D, Ballance T, Campbell C, Dahl E~D, Crawford O, Blunt
  N~S, Rogers B, Noel T and Saffman M 2022
  \href{http://dx.doi.org/10.1038/s41586-022-04603-6}{ {\em Nature\/} {\bf 604}
  457 } \urlprefix\url{https://doi.org/10.1038/s41586-022-04603-6}

\bibitem{Bluvstein2022}
Bluvstein D, Levine H, Semeghini G, Wang T~T, Ebadi S, Kalinowski M, Keesling
  A, Maskara N, Pichler H, Greiner M, Vuleti{\'{c}} V and Lukin M~D 2022
  \href{http://dx.doi.org/10.1038/s41586-022-04592-6}{ {\em Nature\/} {\bf 604}
  451 } \urlprefix\url{https://doi.org/10.1038/s41586-022-04592-6}

\bibitem{Raab1987}
Raab E~L, Prentiss M, Cable A, Chu S and Pritchard D~E 1987
  \href{http://dx.doi.org/10.1103/PhysRevLett.59.2631}{ {\em Physical Review
  Letters\/} {\bf 59} 2631 }

\bibitem{Rushton2014}
Rushton J~A, Aldous M and Himsworth M~D 2014
  \href{http://dx.doi.org/10.1063/1.4904066}{ {\em Review of Scientific
  Instruments\/} {\bf 85} 121501 }
  \urlprefix\url{https://doi.org/10.1063/1.4904066}

\bibitem{McGilligan2022}
McGilligan J~P, Gallacher K, Griffin P~F, Paul D~J, Arnold A~S and Riis E 2022
  \href{http://dx.doi.org/10.1063/5.0101628}{ {\em Review of Scientific
  Instruments\/} {\bf 93} 091101 }
  \urlprefix\url{https://doi.org/10.1063/5.0101628}

\bibitem{Reichel1999}
Reichel J, H{\"{a}}nsel W and H{\"{a}}nsch T~W 1999
  \href{http://dx.doi.org/10.1103/PhysRevLett.83.3398}{ {\em Phys. Rev.
  Lett.\/} {\bf 83} 3398 }

\bibitem{Lee1996}
Lee K~I, Kim J~A, Noh H~R and Jhe W 1996
  \href{http://dx.doi.org/10.1364/OL.21.001177}{ {\em Optics Letters\/} {\bf
  21} 1177 }

\bibitem{Shimizu1991}
Shimizu F, Shimizu K and Takuma H 1991
  \href{http://dx.doi.org/10.1364/OL.16.000339}{ {\em Optics Letters\/} {\bf
  16} 339 }
  \urlprefix\url{https://www.osapublishing.org/abstract.cfm?URI=ol-16-5-339}

\bibitem{Vangeleyn2009}
Vangeleyn M, Griffin P~F, Riis E and Arnold A~S 2009
  \href{http://dx.doi.org/10.1364/OE.17.013601}{ {\em Optics Express\/} {\bf
  17} 13601 } \urlprefix\url{https://doi.org/10.1364/OE.17.013601}

\bibitem{Isichenko2023}
Isichenko A, Chauhan N, Bose D, Wang J, Kunz P~D and Blumenthal D~J 2023
  \href{http://dx.doi.org/10.1038/s41467-023-38818-6}{ {\em Nature
  Communications\/} {\bf 14} 3080 }
  \urlprefix\url{https://doi.org/10.1038/s41467-023-38818-6}

\bibitem{Ropp2023}
Ropp C, Zhu W, Yulaev A, Westly D, Simelgor G, Rakholia A, Lunden W, Sheredy D,
  Boyd M~M, Papp S, Agrawal A and Aksyuk V 2023
  \href{http://dx.doi.org/10.1038/s41377-023-01081-x}{ {\em Light: Science {\&}
  Applications\/} {\bf 12} 83 }
  \urlprefix\url{https://doi.org/10.1038/s41377-023-01081-x}

\bibitem{Vangeleyn2010}
Vangeleyn M, Griffin P~F, Riis E and Arnold A~S 2010
  \href{http://dx.doi.org/10.1364/OL.35.003453}{ {\em Optics Letters\/} {\bf
  35} 3453 } \urlprefix\url{https://dx.doi.org/10.1364/OL.35.003453}

\bibitem{Nshii2013}
Nshii C~C, Vangeleyn M, Cotter J~P, Griffin P~F, Hinds E~A, Ironside C~N, See
  P, Sinclair A~G, Riis E and Arnold A~S 2013
  \href{http://dx.doi.org/10.1038/nnano.2013.47}{ {\em Nature Nanotechnology\/}
  {\bf 8} 321 } \urlprefix\url{www.nature.com/doifinder/10.1038/nnano.2013.47}

\bibitem{Lin1991}
Lin Z, Shimizu K, Zhan M, Shimizu F and Takuma H 1991 {\em Japanese Journal of
  Applied Physics\/} {\bf 30} L 1324

\bibitem{McGehee2021}
McGehee W~R, Zhu W, Barker D~S, Westly D, Yulaev A, Klimov N, Agrawal A, Eckel
  S, Aksyuk V and McClelland J~J 2021
  \href{http://dx.doi.org/10.1088/1367-2630/abdce3}{ {\em New Journal of
  Physics\/} {\bf 23} 013021 }

\bibitem{McGilligan2020}
McGilligan J~P, Moore K~R, Dellis A, Martinez G~D, {De Clercq} E, Griffin P~F,
  Arnold A~S, Riis E, Boudot R and Kitching J 2020
  \href{http://dx.doi.org/10.1063/5.0014658}{ {\em Appl. Phys. Lett.\/} {\bf
  117} 054001 }

\bibitem{Elvin2019}
Elvin R, Hoth G~W, Wright M, Lewis B, McGilligan J~P, Arnold A~S, Griffin P~F
  and Riis E 2019 \href{http://dx.doi.org/10.1364/oe.378632}{ {\em Optics
  Express\/} {\bf 27} 38359 } \urlprefix\url{https://doi.org/10.1364/oe.378632}

\bibitem{Lee2022}
Lee J, Ding R, Christensen J, Rosenthal R~R, Ison A, Gillund D~P, Bossert D,
  Fuerschbach K~H, Kindel W, Finnegan P~S, Wendt J~R, Gehl M, Kodigala A,
  McGuinness H, Walker C~A, Kemme S~A, Lentine A, Biedermann G and Schwindt
  P~D~D 2022 \href{http://dx.doi.org/10.1038/s41467-022-31410-4}{ {\em Nature
  Communications\/} {\bf 13} }
  \urlprefix\url{https://doi.org/10.1038/s41467-022-31410-4}

\bibitem{Ehinger2022}
Ehinger L~H, Acharya B~P, Barker D~S, Fedchak J~A, Scherschligt J, Tiesinga E
  and Eckel S 2022 \href{http://dx.doi.org/10.1116/5.0095011}{ {\em {AVS}
  Quantum Science\/} {\bf 4} 034403 }
  \urlprefix\url{https://doi.org/10.1116/5.0095011}

\bibitem{Barker2019}
Barker D~S, Norrgard E~B, Klimov N~N, Fedchak J~A, Scherschligt J and Eckel S
  2019 \href{http://dx.doi.org/10.1103/PhysRevApplied.11.064023}{ {\em Physical
  Review Applied\/} {\bf 11} 064023 }

\bibitem{Sitaram2020}
Sitaram A, Elgee P~K, Campbell G~K, Klimov N~N, Eckel S and Barker D~S 2020
  \href{http://dx.doi.org/10.1063/5.0019551}{ {\em Review of Scientific
  Instruments\/} {\bf 91} 103202 }
  \urlprefix\url{https://aip.scitation.org/doi/10.1063/5.0019551}

\bibitem{Lee2013}
Lee J, Grover J~A, Orozco L~A and Rolston S~L 2013
  \href{http://dx.doi.org/10.1364/JOSAB.30.002869}{ {\em J. Opt. Soc. Am. B\/}
  {\bf 30} 2869--2874 }

\bibitem{McGilligan2015}
McGilligan J~P, Griffin P~F, Riis E and Arnold A~S 2015
  \href{http://dx.doi.org/10.1364/OE.23.008948}{ {\em Opt. Express\/} {\bf 23}
  8948 }
  \urlprefix\url{http://www.osapublishing.org/viewmedia.cfm?uri=oe-23-7-8948{\&}seq=0{\&}html=true}

\bibitem{Bondza2022}
Bondza S, Lisdat C, Kroker S and Leopold T 2022
  \href{http://dx.doi.org/10.1103/physrevapplied.17.044002}{ {\em Physical
  Review Applied\/} {\bf 17} }
  \urlprefix\url{https://doi.org/10.1103/physrevapplied.17.044002}

\bibitem{Imhof2017}
Imhof E, Stuhl B~K, Kasch B, Kroese B, Olson S~E and Squires M~B 2017
  \href{http://dx.doi.org/10.1103/PhysRevA.96.033636}{ {\em Physical Review
  A\/} {\bf 96} 033636 }

\bibitem{Eckel2018}
Eckel S, Barker D~S, Fedchak J~A, Klimov N~N, Norrgard E, Scherschligt J,
  Makrides C and Tiesinga E 2018
  \href{http://dx.doi.org/10.1088/1681-7575/aadbe4}{ {\em Metrologia\/} {\bf
  55} S182 } \urlprefix\url{http://iopscience.iop.org/10.1088/1681-7575/aadbe4}

\bibitem{Barker2022}
Barker D~S, Norrgard E~B, Klimov N~N, Fedchak J~A, Scherschligt J and Eckel S
  2022 \href{http://dx.doi.org/10.1364/oe.444711}{ {\em Opt. Express\/} {\bf
  30} 9959 }

\bibitem{Tarbutt2015}
Tarbutt M~R 2015 \href{http://dx.doi.org/10.1088/1367-2630/17/1/015007}{ {\em
  New J. Phys.\/} {\bf 17} 015007 }

\bibitem{Eckel2022}
Eckel S, Barker D~S, Norrgard E~B and Scherschligt J 2022
  \href{http://dx.doi.org/10.1016/j.cpc.2021.108166}{ {\em Comput. Phys.
  Commun.\/} {\bf 270} 108166 } \urlprefix\url{http://arxiv.org/abs/2011.07979}

\bibitem{Mukaiyama2003}
Mukaiyama T, Katori H, Ido T, Li Y and Kuwata-Gonokami M 2003
  \href{http://dx.doi.org/10.1103/PhysRevLett.90.113002}{ {\em Phys. Rev.
  Lett.\/} {\bf 90}(11) 113002 }
  \urlprefix\url{https://link.aps.org/doi/10.1103/PhysRevLett.90.113002}

\bibitem{Muniz2018}
Muniz J~A, Norcia M~A, Cline J~R~K and Thompson J~K 2018 A robust narrow-line
  magneto-optical trap using adiabatic transfer [\eprint{1806.00838}]

\bibitem{Snigirev2019}
Snigirev S, Park A~J, Heinz A, Bloch I and Blatt S 2019
  \href{http://dx.doi.org/10.1103/PhysRevA.99.063421}{ {\em Phys. Rev. A\/}
  {\bf 99} 063421 } \urlprefix\url{https://doi.org/10.1103/PhysRevA.99.063421}

\bibitem{Boyd2007}
Boyd M~M 2007 {\em {High Precision Spectroscopy of Strontium in an Optical
  Lattice: Towards a New Standard for Frequency and Time}\/} Ph.D. thesis
  University of Colorado, Boulder

\bibitem{DeSalvo2010}
DeSalvo B~J, Yan M, Mickelson P~G, {Martinez de Escobar} Y~N and Killian T~C
  2010 \href{http://dx.doi.org/10.1103/PhysRevLett.105.030402}{ {\em Phys. Rev.
  Lett.\/} {\bf 105} 030402 }
  \urlprefix\url{http://link.aps.org/doi/10.1103/PhysRevLett.105.030402}

\bibitem{Baumann1966}
Baumann M and Wandel G 1966
  \href{http://dx.doi.org/10.1016/0031-9163(66)90614-7}{ {\em Phys. Lett.\/}
  {\bf 22} 283 }

\bibitem{Kluge1974}
Kluge H~J and Sauter H 1974 {\em Z. Phys.\/} {\bf 270} 295

\bibitem{Berends1992}
Berends R~W and Maleki L 1992 \href{http://dx.doi.org/10.1364/josab.9.000332}{
  {\em J. Opt. Soc. Am. B\/} {\bf 9} 332 }

\bibitem{Nagel2005}
Nagel S~B, Mickelson P~G, Saenz A~D, Martinez Y~N, Chen Y~C, Killian T~C,
  Pellegrini P and Cote R 2005
  \href{http://dx.doi.org/10.1103/PhysRevLett.94.083004}{ {\em Phys. Rev.
  Lett.\/} {\bf 94} 083004 }

\bibitem{Yasuda2006}
Yasuda M, Kishimoto T, Takamoto M and Katori H 2006
  \href{http://dx.doi.org/10.1103/PhysRevA.73.011403}{ {\em Phys. Rev. A\/}
  {\bf 73} 011403(R) }

\bibitem{Kleinert2016}
Kleinert M, {Gold Dahl} M~E and Bergeson S 2016
  \href{http://dx.doi.org/10.1103/PhysRevA.94.052511}{ {\em Phys. Rev. A\/}
  {\bf 94} 052511 }

\bibitem{Foot2005}
Foot C 2005 {\em Atomic Physics\/} Oxford Master Series in Physics (OUP Oxford)
  ISBN 9780198506959

\bibitem{Maruyama2003}
Maruyama R, Wynar R~H, Romalis M~V, Andalkar A, Swallows M~D, Pearson C~E and
  Fortson E~N 2003 \href{http://dx.doi.org/10.1103/PhysRevA.68.011403}{ {\em
  Phys. Rev. A\/} {\bf 68} 011403(R) }

\bibitem{Xu2003a}
Xu X, Loftus T~H, Hall J~L, Gallagher A and Ye J 2003
  \href{http://dx.doi.org/10.1364/JOSAB.20.000968}{ {\em J. Opt. Soc. Am. B\/}
  {\bf 20} 968 }

\bibitem{Tarbutt2015a}
Tarbutt M~R and Steimle T~C 2015
  \href{http://dx.doi.org/10.1103/PhysRevA.92.053401}{ {\em Phys. Rev. A\/}
  {\bf 92}(5) 053401 }

\bibitem{Flemming1997}
Flemming J, Tuboy A, Milori D, Marcassa L, Zilio S and Bagnato V 1997
  \href{http://dx.doi.org/10.1016/S0030-4018(96)00660-8}{ {\em Optics
  Communications\/} {\bf 135} 269--272 }

\end{thebibliography}

\end{document}